\documentclass[prd,aps,superscriptaddress,showpacs,preprintnumbers,amsmath,amssymb,floatfix,12pt]{revtex4}

\usepackage{graphicx}
\usepackage{feynmp}
\unitlength=1.2mm \preprint{BNL-HET-05/11, CU-TP-1120, RBRC-494,}
\preprint{FERMILAB-PUB-05-062-T }
\RequirePackage{feynarts}
\newcommand{\epe}{\epsilon^\prime/\epsilon}
\newcommand{\bea}{\begin{eqnarray}}
\newcommand{\eea}{\end{eqnarray}}

\begin{document}
\bibliographystyle{apsrev}

\title{Systematic effects of the quenched approximation on the strong penguin contribution to
$\epsilon '/\epsilon$}

\author{C.~Aubin}
\affiliation{Physics Department, Columbia University, New York, NY
10027}

\author{N.~H.~Christ}
\affiliation{Physics Department, Columbia University, New York, NY
10027}

\author{C.~Dawson}
\affiliation{RIKEN-BNL Research Center, Brookhaven National Laboratory,Upton, NY 11973}

\author{J.W.~Laiho}
\affiliation{Physics Department, Brookhaven National
Laboratory,Upton, NY 11973} \affiliation{Physics Department,
Princeton University, Princeton, NJ 08544} \affiliation{Present
address: Physics Department, Fermilab, Batavia, IL 60510}

\author{J.~Noaki}
\affiliation{RIKEN-BNL Research Center, Brookhaven National Laboratory,Upton, NY 11973}
\affiliation{Present address: School of Physics and Astronomy, University of
 Southampton, Southampton, SO17 1BJ, England}

\author{S.~Li}
\affiliation{Physics Department, Columbia University, New York, NY
10027}

\author{A.~Soni}
\affiliation{Physics Department, Brookhaven National Laboratory,Upton, NY 11973}


\bibliographystyle{apsrev}

\begin{abstract}
We discuss the implementation and properties of the quenched
approximation in the calculation of the left-right, strong penguin
contributions ({\it i.e.} $Q_6$) to $\epsilon'/\epsilon$.  The
coefficient of the new chiral logarithm, discovered by Golterman
and Pallante, which appears at leading order in quenched chiral
perturbation theory is evaluated using both the method proposed by
those authors and by an improved approach which is free of power
divergent corrections.  The result implies a large quenching
artifact in the contribution of $Q_6$ to $\epsilon'/\epsilon$.
This failure of the quenched approximation affects only the strong
penguin operators and so does not affect the $Q_8$ contribution to
$\epsilon'/\epsilon$ nor $\textrm{Re} A_0$, $\textrm{Re} A_2$ and
thus, the $\Delta I=1/2$ rule at tree level in chiral perturbation
theory.
\end{abstract}

\maketitle

\newpage

\section{Introduction}
\label{sec:introduction}

There have been several recent applications of lattice QCD to the
calculation of $\textrm{Re}(\epe)$, the direct \emph{CP} violating
parameters in $K \to \pi\pi$ decays.  These include the attempts
using domain wall fermions by the CP-PACS \cite{Noaki:2001un} and
RBC \cite{Blum:2001xb} collaborations.  A notable feature of both
of these calculations is that their central values differ
drastically from experiment. The experiments at CERN
\cite{Fanti:1999nm,Batley:2002gn} and Fermilab
\cite{Alavi-Harati:1999xp,Alavi-Harati:2002ye} have yielded an
experimental world average of $\textrm{Re}(\epe) = (1.8\pm
0.4)\times 10^{-3}$ \cite{Eidelman:2004wy}.  The RBC collaboration
reported a value $ -4.0(2.3) \times 10^{-4}$, and a similar
negative central value was reported by CP-PACS.  (See
\cite{Pekurovsky:1998jd} for an earlier attempt using staggered
fermions and \cite{Bhattacharya:2004qu} for ongoing work also
using staggered fermions.) The stated errors in the RBC lattice
calculations for $\epe$ were statistical only, with an estimate of
the size of the systematic errors requiring much further study, of
which this work is part.

A number of serious approximations were made in the lattice
calculations which introduced uncontrolled systematic errors.  One
of these was the quenched approximation, in which the fermion
determinant is ignored in the generation of the gauge
configurations. This is a truncation of the full theory that
dramatically reduces the computer resources required, but is
uncontrolled. Where quenched lattice results have been compared to
experiment, in simple quantities such as masses of flavored mesons
and decay constants, the agreement is at or better than $\sim
15\%$. However, there is no apparent reason for this level of
agreement to hold for all low-energy hadronic phenomena. Another
approximation made in all existing lattice calculations of $\epe$
was the use of leading order chiral perturbation theory (ChPT) to
relate unphysical $K \to \pi$ and $K \to 0$ amplitudes to the
physical $K \to \pi\pi$ amplitudes, as first proposed by
\cite{Bernard:1985wf}. This is also likely to be a serious source
of systematic error, although in this paper we focus on a
particular ambiguity present in the quenched approximation.

Since the original lattice calculations of
\cite{Blum:2001xb,Noaki:2001un}, it was shown in
\cite{Golterman:2001qj,Golterman:2002us} that, at leading order in
quenched chiral perturbation theory there is a term logarithmic in
the pion mass which contributes to the matrix elements of the
strong penguin operators.  This term is absent in full QCD, and
its contribution is proportional to an \emph{a priori} unknown,
new low energy constant (LEC).  In terms of a representation of
quenched QCD in which the fermion loops are cancelled by the
addition of ghost fields to the lagrangian, this LEC can be
associated with the presence of additional operators in the
effective hamiltonian mediating the weak decay of $K\to\pi\pi$
which contain both quark and ghost fields.  The presence of
additional operators naturally calls for a re-examination of the
way in which the quenched approximation is implemented in such
matrix element calculations; without further guidance, a single
strong penguin operator in the full theory can be represented by
any arbitrary linear combination of its direct transcription into
the quenched theory (a four-quark operator) and these additional
(two-quark, two-ghost) operators.



Since this ChPT result only became known after the original RBC
analysis was completed \cite{Blum:2001xb}, that work used the
leading order ChPT relevant for full QCD.  As such, it is useful
to discuss how the presence of these new operators might effect
this result.  We emphasize that this particular quenching
difficulty is present only for the strong penguin operators and so
does not affect the $Q_8$ contribution to $\epe$. This quenching
ambiguity also does not significantly affect Re $A_0$ and Re $A_2$
and since Re $A_0$ receives only a negligibly small contribution
from $Q_6$ \cite{Blum:2001xb}, the RBC result for the $\Delta
I=1/2$ rule remains unchanged.  A recent paper by Golterman and
Pallante \cite{Golterman:2006ed} shows how similar quenched
penguin effects can arise in the $\Delta I = 1/2$ rule, and thus
change the values of the underlying LEC's that one measures.
However, the effects of Ref.~\cite{Golterman:2006ed} do not change
the tree-level results of Bernard, et al \cite{Bernard:1985wf},
and as the original RBC analysis was performed at tree level, this
does not affect the actual results for the $\Delta I = 1/2$ rule.
An attempt to determine the systematic error due to quenching for
this quantity may, however, be influenced by the results of
Golterman and Pallante.

Since the quenched approximation is uncontrolled, a rigorous
matching of operators between the quenched and full theories is
not possible. However, we argue that the coefficients of these new
operators can be determined by the same style of physical argument
that is usually put forward to motivate the quenched
approximation.  This approach, which might be called
``intermediate-energy matching'' can be described as follows.
Since quark loops play an important role in the renormalization
group evolution of the weak amplitudes from the scale of the $W$,
$Z$ and top quark down to the kaon mass, the quenched
approximation must be applied in a discriminating fashion,
simulating the vacuum polarization of quark loops in the
low-energy lattice QCD calculation with a weakened bare coupling
while leaving the quark loops that appear in the perturbatively
computed Wilson coefficients intact.

Such a separation between short- and long-distance vacuum
polarization effects can be made quite precise in the context of
the effective weak Hamiltonian and gives a prescription for
carrying out the quenched calculation which is close to that
adopted by RBC and CP-PACS.  One simply requires at the
intermediate energy scale at which the perturbatively determined
weak amplitudes are matched to the low-energy four-quark operators
in $H_W$ that the full- and quenched-QCD amplitudes agree.  Since
there are no ghost quarks in the full theory, this requires that
the ghost quark matrix elements of $H_W$ evaluated in the quenched
theory also vanish.  This gives a physically motivated definition
of the quenched approximation for the energy scale at which this
matching is performed.  Of course, as the quenched and full
theories do not have the same low energy limit, physical results
will depend on this matching scale; for the quenched approximation
to be useful, the difference in these results as the matching
scale is varied over the range of energy scales important for the
quantities we are calculating must be numerically small.  The
extent to which this condition is obeyed depends on the quantity
calculated; later we argue that for the strong penguin operators,
the quenched approximation is particularly bad.

Another possible guide we may take in transcribing the full QCD
operators into the quenched theory is the extended chiral symmetry
of this theory~\cite{Golterman:2001qj,Golterman:2002us}. Recall
that when the ghost quarks are added to normal QCD, the original
chiral symmetry, $SU_L(3)\times SU_R(3)$, which transforms only
the normal quarks, expands into a larger, graded symmetry,
$SU_L(3|3)\times SU_R(3|3)$ for the simplest case of three quarks
$u$, $d$ and $s$ and three ghost quarks $\tilde{u}$, $\tilde{d}$
and $\tilde{s}$.

As analyzed by Golterman and Pallante, the original operators appearing
in $H_W$, transforming in specific representations of
$SU_L(3)\times SU_R(3)$, take on new $SU_L(3|3)\times SU_R(3|3)$ quantum
numbers.  The new operators containing ghost quarks might then be
chosen in a fashion to simplify the resulting representation of the
extended $SU_L(3|3)\times SU_R(3|3)$ symmetry group while leaving the
physical $SU_L(3)\times SU_R(3)$ behavior unchanged.  However, since
this extended $SU_L(3|3)\times SU_R(3|3)$ group is necessarily
unphysical it is difficult to provide a convincing motivation for the
mixture of ghost quarks chosen.

Thus, we believe that the quenched approximation can be applied to
weak decays in a well-motivated way.  One determines by
``intermediate-energy matching", the quenched effective
Hamiltonian, $H_W^{qh}$, and then examines matrix elements of the
resulting operators analytically for possible quenched chiral
logarithms and numerically to make quenched predictions from the
theory.  The size of the quenched chiral logarithms should be
viewed as a measure of the errors in the quenched approximation.

From this perspective, the study of the quenched chiral logarithm
discovered by Golterman and Pallante and the new low energy constant,
$\alpha_q^{NS}$, which appears as its coefficient, is of central importance.
In fact, these authors have proposed~\cite{Golterman:2002us} a method for obtaining
$\alpha_q^{NS}$, directly from a lattice calculation.   We show that there
are difficulties using their method to obtain $\alpha_q^{NS}$ from lattice
data using domain wall fermions, due to the presence of power divergences.
However, motivated by their direct approach, we have found an improved
method for obtaining $\alpha_q^{NS}$ that does not have these divergences.
This is done by constructing an extension of the usual CPS symmetry arguments
to the quenched case where both quark and ``ghost" quark degrees of freedom
are present.

Using our proposed method, we obtain a value of $\alpha_q^{NS}$
which is consistent with the value given by the large $N_{c}$
approximation obtained by Golterman and Peris
\cite{Golterman:2003uj, Golterman:2003yw}. The value obtained from
the method proposed by Golterman and Pallante also yielded a value
roughly in agreement with the others, but with a larger systematic
error. This large value of $\alpha_q^{NS}$ has two important
consequences:  i) Large, non-analytic behavior in the quenched
chiral limit which is absent in full QCD provides clear evidence
of substantial systematic errors associated with the quenched
approximation. ii) If these new quenched non-linearities are
omitted from the functional forms used to extract the quenched
LEC's, even the analysis within the quenched approximation is
likely to be incorrect.

In Section~\ref{sec:quenching} we discuss the application of the
quenched approximation to weak decay amplitudes and motivate the
``intermediate-energy matching'' approach described above. The
chiral symmetry of partially quenched QCD and quenched chiral
perturbation theory is reviewed in Sec.~\ref{sec:quenched_chpt}.
We follow the approach of Ref~\cite{Golterman:2001qj} and discuss
the ambiguities of determining $H_W^{qh}$ from this perspective.
In Section~\ref{sec:alpha_NS_K_vac_thry} we discuss Golterman and
Pallante's method~\cite{Golterman:2002us} for obtaining the low
energy constant, $\alpha_q^{NS}$.  We extend the usual CPS
symmetry arguments for obtaining the form of the power divergent
contributions to $\Delta I=1/2$ matrix elements to the quenched
case involving  external ghost quark states.  In
Section~\ref{sec:alpha_NS_K_vac_num} we use this extended
symmetry, as well as numerical results, to show that the method of
\cite{Golterman:2002us}, when evaluated using domain wall
fermions, suffers from ambiguities due to power divergent
contributions.

In Section~\ref{sec:alpha_NS_K_pi_thry} an alternative method for
obtaining $\alpha_q^{NS}$ is proposed which does not have a
divergent contribution and this method is used in
Section~\ref{sec:alpha_NS_K_pi_num} to obtain a numerical value
for this constant. Finally, the implications of this result are
discussed in the conclusion, Section~\ref{sec:conclusion}. Some
useful formulae are given in Appendix A, our conventions are
specified in detail in Appendix B, and a discussion of the
one-loop ChPT calculation used in the numerical extrapolation to
obtain $\alpha_q^{NS}$ is given in Appendix C.

\section{Quenched approximation in weak decay amplitudes}
\label{sec:quenching}

While the quenched approximation is widely used as a device to
reduce the computational requirements of lattice QCD calculations,
its implementation in the calculation of weak matrix elements
deserves further discussion. The physical justification for this
approximation is the hypothesis that the largest effect of the
omitted quark loops is a modification of the running QCD coupling
constant $\alpha_s(\mu)$ because of the omission of quark vacuum
polarization.  To the extent that this hypothesis holds true for a
particular quantity of physical interest we can then compensate to
a large extent for the effects of quenching on $\alpha_s(\mu)$ at
a relevant physical scale $\mu$ by an appropriate weakening of the
bare lattice coupling $\alpha_0 = g^2_0/4\pi$ where $g_0$ is the
gauge coupling which appears directly in the lattice QCD
Lagrangian.  An important limitation of this justification is
apparent from the scale dependence of $\alpha_s(\mu)$.  This scale
dependence will be different between the full and quenched
theories so the equality $\alpha_s(\mu) = \alpha_s^{qh}(\mu)$ will
be approximately true only over a limited range of scales $\mu$.

Since a weak matrix element involves a large range of energy scales from the
top quark mass down to $\Lambda_{\rm QCD}$, a naive application of the quenched
approximation to this entire energy range would be very inaccurate.  However,
because perturbation theory is used to describe most of this larger energy
range, it is also not necessary to use the quenched approximation over
the entire range.  A clear view of this situation comes from considering
some sample Feynman graphs representing the various ways that quark vacuum
polarization loops can enter the gluonic penguin diagrams of interest.

The possible effects of quark vacuum polarization are illustrated
in Figs.~\ref{fig:vac_pol_sd}-\ref{fig:vac_pol_sd_and_ld}.  In the
first of these, Fig.~\ref{fig:vac_pol_sd}, the quark loop is
entirely contained in the short distance part of the graph.  The
vacuum polarization loop will be dominated by momenta on the order
of the top mass and can be accurately treated in QCD perturbation
theory. Contributions with loop momentum on the order of
$\Lambda_{\rm QCD}$, where perturbation theory would not be
accurate, will be suppressed by a factor of $(\Lambda_{\rm
QCD}/m_{top})^2$

By contrast, the diagram shown in Fig.~\ref{fig:vac_pol_ld}
contains a vacuum polarization loop that can only contain low
momentum. More precisely, once multiplicative charge and
wave-function renormalization constants have been removed, any
further contributions from large energies, for example $\mu
\approx m_W$, will be suppressed by an additional factor of
$\mu^2/m_W^2$. Thus this graph represents a potentially
non-perturbative piece which requires lattice techniques to
evaluate.  In fact, the above quenching hypothesis is simply the
statement that the most important effect of removing this quark
loop is a change in the charge renormalization, a change that can
be completely compensated by a corresponding change in the bare
lattice coupling $g_0^2$.

The diagram shown in Fig.~\ref{fig:vac_pol_sd_or_ld} contains a vacuum
polarization loop that can either enter a short- or long-distance part
of the graph.  These two possibilities are represented by the two
dotted boxes shown in the figure.  Because of the presence of the $W$
propagator, the inner box is necessarily a ``high-momentum'' subgraph.
However, important contributions can come from regions of momentum
space in which the lines entering the vacuum polarization subgraph
carry either small ($\approx \Lambda_{\rm QCD}$) or large ($\approx m_W$)
momentum.  In the case that large momentum is involved, the outer dotted
box surrounds what becomes a ``high-momentum'' subgraph with all internal
lines carrying large momentum: a regime that can be evaluated
perturbatively and one in which the quenched approximation should not be
used.  For the ``low-momentum'' case, the contribution will be
non-perturbative and evaluation using lattice techniques, potentially
using the quenched approximation, will be needed.  Other momentum assignments
for the lines in this outer box which include only a portion of the vacuum
polarization subgraph will have more than four external quark lines, thereby
having larger dimension and giving a contribution that is suppressed by a
factor of $(\Lambda_{\rm QCD}/m_W)^2$.

Finally, Fig.~\ref{fig:vac_pol_sd_and_ld} shows a potentially more
ambiguous case.  Here, in addition to the possibility that the
vacuum polarization loop is contained entirely within a high- or
low-momentum subgraph, (the outer and inner dotted boxes
respectively) it may also be partially in both, as indicated by
the middle dotted box.  It is this intermediate subgraph which
requires special discussion. For the case that all the momentum in
the subgraph contained in the outer dotted box is large, the
vacuum polarization loop can be treated perturbatively without
recourse to the quenched approximation.  For the case that only
the momentum contained in the inner loop is large and those in the
remainder of the graph are small, non-perturbative techniques will
be needed and the quenched approximation may be used.  Thus, in
this case the vacuum polarization loop may be removed, or
equivalently, a cancelling loop of ghost quarks included.

However, for the case of the intermediate dotted box which cuts
the quark loop, the vacuum polarization piece is one-half within
the high momentum part (the contents of this intermediate dotted
box) but one half is to be evaluated in the low-momentum part.
Here it is less obvious whether this loop is to be included
perturbatively ({\it i.e.} incorporated in a Wilson coefficient)
or to be cancelled by an added ghost quark contribution.  In the
language of the effective weak Hamiltonian $H_W$, such a
cancellation of this vacuum polarization graph would be achieved
by including ghost operators in $H_W$.

This variety of roles played by QCD vacuum polarization graphs might
suggest that a precise definition of the quenched approximation for
the evaluation of weak matrix elements would require new and elaborate
development.  However, as the above examples suggest, the standard
field theoretic formulation of the ``effective'' low energy theory nicely
deals with all of these questions, providing an unambiguous separation
of the decay amplitudes into short-distance perturbative parts in
which no quenched approximation is made and long-distance parts that
must be evaluated non-perturbatively, possibly within a quenched
approximation.  The potential ambiguity associated with the inclusion
of ghost operators is resolved by the matching conditions
that are imposed to define the quenched effective theory.

To be more concrete, consider the gluonic penguin portion of the
effective low-energy weak Hamiltonian transforming in the $(8,1)$
representation of $SU(3)_L \times SU(3)_R$ and written as a sum
of four independent, dimension-six, four-quark operators:
\begin{equation}
H_W = c_3 Q_3 + c_4 Q_4 + c_5 Q_5 + c_6 Q_6 \label{eq:eff_hw}
\end{equation}
where $\{c_i\}_{i=3-6}$ are the four Wilson coefficients and
$\{Q_i\}_{i=3-6}$ the four, conventional gluonic penguin
operators:
\begin{eqnarray}
Q_3 &=&  \sum_{q=u,d,s} \overline{s}_a \gamma^\mu (1-\gamma^5) d_a
                        \overline{q}_b  \gamma^\mu (1-\gamma^5) q_b
                        \label{eq:o3}\\
Q_4 &=&  \sum_{q=u,d,s} \overline{s}_a \gamma^\mu (1-\gamma^5) d_b
                        \overline{q}_b  \gamma^\mu (1-\gamma^5) q_a
                        \label{eq:o4}\\
Q_5 &=&  \sum_{q=u,d,s} \overline{s}_a \gamma^\mu (1-\gamma^5) d_a
                        \overline{q}_b  \gamma^\mu (1+\gamma^5) q_b
                        \label{eq:o5}\\
Q_6 &=&  \sum_{q=u,d,s} \overline{s}_a \gamma^\mu (1-\gamma^5) d_b
                        \overline{q}_b  \gamma^\mu (1+\gamma^5) q_a.
                        \label{eq:o6_1}
\end{eqnarray}
A sum over the repeated color indices $a$ and $b$ is understood.

The above discussion of the contributions of the various vacuum
polarization graphs to short- and long-distance physics mirrors
closely the usual field-theoretic derivation of the effective weak
Hamiltonian in Eq.~\ref{eq:eff_hw}.  The usual ``factorization''
of weak amplitudes into short- and long-distance parts realized by
$H_W$ provides exactly the needed separation of vacuum
polarization effects into those that are correctly included using
perturbation theory and those which are omitted in the quenched
approximation---their effects being partially reproduced by a
decrease in the bare coupling constant.

Let us briefly recall this correspondence.  The matrix elements of the
local effective weak Hamiltonian given in Eq.~\ref{eq:eff_hw} will accurately
reproduce those of the complete theory if those matrix elements involve momenta
which are small compared to the scale of the $W$ meson and top quark masses---the
scale at which non-trivial structure for the weak interactions becomes visible.

Leading contributions in an expansion in $1/m_{top}^2$ and
$1/m_W^2$ will come from regions of integration over internal
momenta in which: a) All momenta in a particular subgraph are
large. b) That subgraph contains any top quark and $W$ boson
internal lines, and c) That subgraph has itself the minimum number
of external lines or, more precisely, represents the lowest
possible mass dimension or largest possible degree of divergence.
Under these circumstances, this subgraph can be treated as a
structureless local composite operator made up of fields
corresponding to the external lines of the subgraph. When
evaluated, integration over this large-momentum region for the
subgraph contributes to the coefficient $c_i$ appearing in $H_W$.

In fact, the four coefficients $\{c_i\}_{i=3-6}$ can be simply
defined as those required to make the complete and effective
theories agree.  Typically the Wilson operators $\{Q_i\}_{i=3-6}$
will be defined by normalization conditions specified at an energy
scale $\mu$ making both the operators $Q_i$ and coefficient
functions $c_i$ functions of this scale $\mu$ as well.  Usually
the scale $\mu$ is also the scale at which the $c_i$ are
determined by requiring specific Greens functions containing an
$H_W$ vertex to agree with those predicted by the complete theory.

The application of this effective field theory formalism to the definition
of the quenched approximation for the evaluation of weak matrix elements is
now quite straight-forward.  The vacuum polarization graphs which contribute
to the short-distance/high-momentum part of the above analysis will necessarily
contribute to the Wilson coefficients, can be evaluated in perturbation
theory and need not involve the quenched approximation.  Vacuum polarization
effects which involve low momentum will enter the matrix elements of the Wilson
operators $Q_i$, will likely require non-perturbative techniques for evaluation
and may be computed in the quenched approximation.

Now the equality between matrix elements of the complete and the
effective theories, imposed for four specific amplitudes at the
scale $\mu$ will hold for a range of scales and a variety of
matrix elements only to the extent that the quenched approximation
is accurate and the matrix elements (e.g. $\langle \pi\pi|Q_i|K
\rangle$) of the $Q_i$ in the unquenched theory agree with those
evaluated in the quenched theory with the bare coupling weakened
to compensate for the omitted quark-anti-quark screening.  Note,
in this definition of the quenched approximation we are replacing
the matrix elements of the operators $Q_i$, originally to be
evaluated in full QCD, with matrix elements evaluated in a new
theory: a theory in which ghost quarks have been introduced,
following the procedures of Bernard and
Golterman~\cite{Bernard:1992mk}, to completely cancel the fermion
determinant, and the gauge coupling at short distances has been
decreased to compensate for this quark loop omission.

Finally, let us examine the potential ambiguity associated with
vacuum polarization loops, such as those in
Fig.~\ref{fig:vac_pol_sd_and_ld}, which contribute partially to
the short-distance, perturbative Wilson coefficients and partially
to the low-energy non-perturbative matrix elements.   The
``quenching'' of the loop cut by the middle dotted box in
Fig.~\ref{fig:vac_pol_sd_and_ld} reduces to the question of
whether we add to our four Wilson operators $\{Q_i\}_{i=3-6}$ of
Eqs.~\ref{eq:o3}-\ref{eq:o6_1} new operators which contain ghost
quarks.  In fact, in the quenched theory there are four additional
operators which have the same symmetry under $SU(3)_L \times
SU(3)_R$ as these original four operators:
\begin{eqnarray}
Q^*_3 &=&  \sum_{\tilde{q}=\tilde{u},\tilde{d},\tilde{s}}
               \overline{s}_a \gamma^\mu (1-\gamma^5) d_a
               \overline{\tilde{q}}_b  \gamma^\mu (1-\gamma^5) \tilde{q}_b
                        \label{eq:o*3}\\
Q^*_4 &=&  \sum_{\tilde{q}=\tilde{u},\tilde{d},\tilde{s}}
               \overline{s}_a \gamma^\mu (1-\gamma^5) d_b
               \overline{\tilde{q}}_a  \gamma^\mu (1-\gamma^5) \tilde{q}_b
                        \label{eq:o*4}\\
Q^*_5 &=&  \sum_{\tilde{q}=\tilde{u},\tilde{d},\tilde{s}}
               \overline{s}_a \gamma^\mu (1-\gamma^5) d_a
               \overline{\tilde{q}}_b  \gamma^\mu (1+\gamma^5) \tilde{q}_b
                        \label{eq:o*5}\\
Q^*_6 &=&  \sum_{\tilde{q}=\tilde{u},\tilde{d},\tilde{s}}
               \overline{s}_a \gamma^\mu (1-\gamma^5) d_b
               \overline{\tilde{q}}_b  \gamma^\mu (1+\gamma^5) \tilde{q}_a
                        \label{eq:o*6}.
\end{eqnarray}
Including these operators in our quenched effective weak Hamiltonian will have
the effect of introducing ghost quark loops which will (partially) cancel the
problematic quark loop in Fig.~\ref{fig:vac_pol_sd_and_ld}.

However, given the discussion above, the most consistent approach to the
quenched approximation is now clear.  The effective, quenched weak Hamiltonian,
$H^{qh}_w$ should incorporate all eight possible operators:
\begin{equation}
H^{qh}_w = \sum_{i=3}^6 \Bigl\{c_i Q_i + c^*_i Q^*_i\Bigr\}
\label{eq:eff_hw_qh}
\end{equation}
and the eight coefficients $\{c_i\}_{i=3-6}$ and
$\{c^*_i\}_{i=3-6}$ should be chosen to make the matrix elements
of $H^{qh}_w$ and those of the complete theory agree as closely as
possible.  This ``matching'' condition should be imposed at a
sufficiently large scale $\mu$ that the required, perturbative
evaluation of the complete theory is justified.  On the other
hand, given the inconsistent $\mu$ dependence of quantities in the
complete and quenched theories, the scale $\mu$ also should be
chosen as close as possible to the low energy region in which the
quenched matrix elements are to be evaluated to minimize these
inconsistencies. The scale $\mu$ is thus an intermediate energy
scale and this approach might be called intermediate energy
matching.

How are these four new coefficients $\{c^*_i\}_{i=3,4,5,6}$ to be evaluated and what
values are expected?  These new coefficients can be fixed by imposing conditions
on the new ghost-quark amplitudes that appear in the quenched theory.  They can
be evaluated by comparing the complete theory, evaluated perturbatively, with
the quenched theory evaluated either perturbatively or non-perturbatively.
Since the complete theory does not contain ghost quarks, we should require that
specific two-quark/two-ghost-quark amplitudes should vanish in the quenched
theory.  This would appropriately be done at off-shell, non-exceptional momentum
at the scale $\mu$ and be imposed on color mixed and unmixed and left- and
right-handed flavor-singlet combinations of ghost quarks.  If carried out in
perturbation theory, the absence of ghost quark coupling in the complete theory
simply requires that all the $c_i^*$ vanish to leading order in $\alpha_s$.  The
choice $c_i^* = 0 |_{i=3,4,5,6}$ is precisely the quenched approximation used
by RBC and CP-PACS in their quenched kaon decay
calculations,~\cite{Noaki:2001un,Blum:2001xb}.  Because of the ghost quark couplings
introduced by the self-contractions of Fig.~\ref{fig:induced_ghosts}, these $c_i^*$
coefficients will be non-zero at order $\alpha_s$.  While it would not be especially
difficult to calculate the $\{c_i^*\}_{i=3,4,5,6}$ to order $\alpha_s$ in perturbation
theory or to evaluate them using the RI-MOM techniques employed for similar quantities
by the RBC collaboration in Ref.~\cite{Blum:2001xb}, we expect that these effects
will be quite small as were similar disconnected amplitudes that were evaluated in
this earlier work.

The majority of numerical simulations and the discussion above has
been focused on the quenched approximation, in which all vacuum
polarization loops are dropped from the calculation.  However,
there is additional insight and a useful framework for analysis to
be obtained in the case of ``partial quenching'' when a portion of
the fermion determinant is included in the calculation.  An
important example involves the use of the wrong number of sea
quarks, working with two rather than three flavors as in
\cite{Aoki:2004ht}. In fact with a positive Dirac operator (such
as with domain wall fermions) and the proper
algorithm~\cite{Clark:2004cp}, one can simulate with a value of
$N_f$ varying continuously between 0 and 3.  The discussion for
the quenched case applies quite directly to this partially
quenched case as well.

Again one attempts to approximate matrix elements computed in the
full theory with those computed in the truncated theory without the
full fermion determinant.  The effects of the missing determinant are
partially reproduced by adjusting the bare coupling to account for
the missing polarization effects.  Again the effective weak Hamiltonian
to be used in a partially quenched theory should be chosen so that at
an intermediate matching point, the full and the partially quenched
theory agree.  Clearly as $N_f \rightarrow 3$ this approximation will
become increasingly accurate as the vacuum polarization effects of the
two theories become identical.

Even in the case $N_f=3$ one can consider a partially quenched
theory in which the valence quark masses do not match with the sea
quark masses in the fermion determinant.  For this case, the above
discussion of quenching is simplified, since the differences
between fermion masses appearing in quark loops can be neglected
at the matching scale $\mu$. Thus, the effective weak Hamiltonian,
written in terms of the valence quark fields, will be the same as
that in the full theory.  Often setting $m_{\rm val} = m_{\rm
sea}$, which can always be done, is less interesting than using
$m_{\rm val} \ne m_{\rm sea}$ together with analytic results to
explore the chiral limit.

In this section we have proposed a definition of the quenched approximation which
provides a natural and self-consistent application of the standard quenched
approximation to the case of hadronic weak decays.  Up to corrections which are
likely small, the calculations of both the RBC and CP-PACS collaborations use this
definition in their evaluations of the gluonic penguin
contributions,~\cite{Noaki:2001un,Blum:2001xb}.   However, since such an approximation
is necessarily {\it ad hoc} and not systematic it must be used with suspicion.
If unphysical effects, such as the quenched chiral logarithms discovered by
Golterman and Pallante and evaluated numerically below, turn out to be large (as we
will see they do), then the quenched approximation to these gluonic penguin
amplitudes should be abandoned.

\section{Review of strong penguins in quenched ChPT}
\label{sec:quenched_chpt}

Following Golterman and Pallante, we will now use chiral symmetry and
chiral perturbation theory to study the effects of the quenched
approximation on the gluonic penguin contributions to $K$ meson decay.
Chiral perturbation theory is an important tool which provides an
approximation scheme in which two-pion decay amplitudes can be computed
from vacuum and single-pion transition amplitudes.  It can also provide
an indication of the accuracy of the quenched approximation by identifying
unphysical, quenched singularities which give rise to the infamous
``quenched chiral logarithms''.

In order to exploit chiral symmetry in the quenched approximation,
we must adopt a field-theoretic description of quenching.  This
can be done in two ways: the supersymmetric formulation
\cite{Bernard:1992mk} and the replica method
\cite{Damgaard:2000gh}.  As in the discussion above, we will adopt
the original supersymmetric formulation and use the quenched
chiral perturbation theory of Bernard and Golterman
\cite{Bernard:1992mk}.  In this method, the valence quarks are
quenched by introducing ghost quarks which have the same mass and
quantum numbers as the valence quarks but opposite statistics.
Therefore the ghost loops cancel the loops of the valence quarks,
effectively setting the fermion determinant to a constant, which
is precisely the quenched approximation.  The chiral symmetry
group of this action is $SU(n|n)_{L} \times SU(n|n)_{R}$, a graded
symmetry group, where $n$ is the number of valence quark flavors.
We adopt the definitions and notation of Bernard and Golterman
\cite{Bernard:1992mk} needed for this supersymmetric approach (see
also Ref.~\cite{DeWitt:1992cy}).

As in Ref.~\cite{Golterman:2000fw}, one can also consider the partially quenched
case, in which $N$ sea quarks are introduced into the theory. In
the notation of \cite{Golterman:2000fw}, one has $n$ quarks, $N$ of which
are sea quarks, so that there are $n-N$ valence and, in addition, $n-N$ ghost
quarks.  The valence quarks have arbitrary mass, while the sea quarks are
chosen to be degenerate. In this case, the graded chiral symmetry
group of the action is $SU(n|n-N)_{L} \times SU(n|n-N)_{R}$.

\subsection{Notation and graded groups}

We will need to identify representations of the graded symmetry
$SU(n|n-N)$ group and will adopt the following notation.  Since
the fundamental representation of $SU(N)$ and its complex
conjugate are usually denoted, $N$ and $\overline{N}$, we will
adopt a similar description of those representations of
$SU(n|n-N)$: $(n|n-N)$ and $\overline{(n|n-N)}$. Thus, for
example, a quark bilinear of the form $\overline{{\cal
Q}}(1+\gamma^5){\cal Q}'$, will belong to a representation easily
identified as  $(\overline{(n|n-N)}_L, (n|n-N)_R )$.  Here we use
${\cal Q}$ to represent a column vector whose first $n$ components
are the anti-commuting quark fields $q$ and whose final $n-N$
components are the ghost quark fields $\tilde{q}$.  Likewise the
quark bilinear $\overline{{\cal Q}}\gamma^\mu(1+\gamma^5){\cal
Q}'$ will belong to the product representation $(1(n|n-N)_L,
\overline{(n|n-N)}_R \times (n|n-N)_R)$ where we use the notation
$1(n|n-N)$ to identify the trivial representation of the group
$SU(n|n-N)$.  This representation is easily constructed as the
$(2n-N) \times (2n-N)$ identity matrix $I^{(n|n-N)}_{a,b}$, where
the indices $a$ and $b$ transform as elements of the $(n|n-N)$ and
$\overline{(n|n-N)}$ representations of $SU(n|n-N)$ respectively.
This matrix has supertrace (defined below in
Eq.~\ref{eq:penguin_qs} and Ref.~\cite{DeWitt:1992cy})
str$(I^{(n|n-N)})=N$.

We will be interested in two irreducible representations which
appear in the product $\overline{(n|n-N)} \times (n|n-N)$ above.
The first is the trivial representation $1(n|n-N)$ already
discussed.  The second is the adjoint representation which we
denote as adj$(n|n-N)$.  This is the representation formed from
the $(2n-N)^2-1$ generators $T^i_{a,b}$ of the group $SU(n|n-N)$.
Here $i$ identifies the element of adj$(n|n-N)$ while the indices
$a$ and $b$ transform as elements of the $(n|n-N)$ and
$\overline{(n|n-N)}$ representations of $SU(n|n-N)$ respectively.
Since the matrices $U$ defining $SU(n|n-N)$ have unit
superdeterminant, sdet$(U)$=1, the generators $T^i$ each have
vanishing supertrace: str$(T^i)=0$.  For the case $N \ne 0$, the
trivial and adjoint representations are entirely distinct.
Operators belonging to $1(n|n-N)$ cannot mix with those in
adj$(n|n-N)$; the latter have a vanishing supertrace while the
former do not.

However, for the quenched case where there are no sea quarks,
$N=0$, their supertraces both vanish and an operator in the
adj$(n|n)$ representation can mix with a $1(n|n)$ operator.  In
fact, in the quenched, $N=0$ case, the multiplication of the $2n$
fields $q$ and $\tilde{q}$ by a common phase factor has
superdeterminant 1 making the unit matrix $I^{(n|n)}_{a,b}$ a
valid generator of $SU(n|n)$ which is no longer simple in this
$N=0$ case.  Thus, for the case $N=0$ the adjoint representation,
adj$(n|n)$, continues to have dimension $(2n)^2 -1$ but now
includes the identity matrix $I^{(n|n)}$.  Of course there are
still anti-hermitian matrices with non-vanishing supertrace.  To
include these we must extend the adjoint representation to a
larger $(2n)^2$-dimensional representation, denoted here
$\overline{\rm adj}(n|n)$ which includes the vectors in
adj$(n|n)$.

Therefore, the product representation $\overline{(n|n)} \times (n|n)$ must
be decomposed in an unfamiliar way.  In contrast with the $N \ne 0$ case,
this representation is not reducible and cannot be written as a direct sum
of adj$(n|n-N)$ and $1(n|n-N)$.  The equation
\begin{equation}
\overline{(n|n-N)} \times (n|n-N) = {\rm adj}(n|n-N) + 1(n|n-N)
\end{equation}
does not apply for the case $N=0$.  Instead this product forms the irreducible
representation $\overline{\rm adj}(n|n)$.  Contained within the
$(2n)^2$-dimensional supervector space on which this representation acts
is an invariant supervector subspace of dimension $(2n)^2-1$ which forms the
representation adj$(n|n)$.  Finally this supervector subspace of dimension
$(2n)^2-1$ itself contains a one-dimensional subspace which is again
invariant under the original $\overline{\rm adj}(n|n)$ representation
matrices, transforming as the trivial $1(n|n)$ representation:
\begin{equation}
\overline{(n|n)} \times (n|n) = \overline{\rm adj}(n|n) \supset
                                    {\rm adj}(n|n) \supset 1(n|n)
\end{equation}
The characteristics of this quenched, $N=0$ case will be discussed further
below.

\subsection{Quenched chiral symmetry of $Q_6$}

Let us now examine the effect that the quenched approximation has
on the chiral symmetry properties of the gluonic penguin operator $Q_6$,
the four-quark operator which is expected, along with $Q_8$, to
make the largest contribution to $\epsilon'/\epsilon$.  Our discussion
builds on that of Golterman and Pallante\cite{Golterman:2001qj,Golterman:2002us}.
Recall that in the effective weak Hamiltonian for the full theory the
$Q_6$ operator is given by Eq.~\ref{eq:o6_1}, repeated here for convenience:
\begin{equation}\label{eq:q6_2}
Q_{6} = \overline{s}_{a} \gamma_{\mu} (1-\gamma^{5}) d_{b} \sum_{q}
        \overline{q}_{b} \gamma^{\mu} (1+\gamma^{5}) q_{a}.
\end{equation}
The right-hand factor in this operator is a sum over light flavors, $q=u, d, s$,
so in the full theory this factor is a flavor singlet under the symmetry group
$SU(3)_R$.  As discussed in the previous section, in the (partially) quenched
theory, vacuum polarization effects permit this operator to mix with a new
operator which contains sea and ghost quarks and belongs to the singlet
representation of $SU(n|n-N)_R$:
\begin{equation}\label{eq:q6s}
Q^S_6  = \overline{s}_{a} \gamma_{\mu} (1-\gamma^{5}) d_{b} \sum_{{\cal Q}}
        \overline{{\cal Q}}_{b}\gamma^{\mu} (1+\gamma^{5}) {{\cal Q}}_{a}.
\end{equation}
where the sum over ${\cal Q}$ contains all valence, sea and ghost
quarks. An appropriate multiple of this operator can be subtracted
from the original $Q_6$ operator, to create a new operator which
transforms in the adjoint representation of $SU(n|n-N)_R$:
\begin{equation}\label{eq:q6adj}
Q^{\rm adj}_6 = Q_6-\frac{3}{N} Q^S_6.
\end{equation}
Here the factor of $3/N$ is easily chosen so that the adjoint operator
$Q^{adj}_6$ has a vanishing supertrace.  Following Golterman and Pallante,
we re-order Eq.~\ref{eq:q6adj} to express the original operator $Q_6$
in terms of $Q^{adj}_6$ and $Q^S_6$:
\begin{equation}\label{eq:q6_3}
Q_6 = Q^{\rm adj}_6 + \frac{3}{N} Q^S_6.
\end{equation}
This is a useful equation because the adjoint representation, to
which $Q^{\rm adj}_6$ belongs, also includes the usual
electro-weak penguin operator $Q_8$ so the matrix elements of
these two operators are connected by a simple supersymmetry
transformation.  In addition, the NLO chiral perturbation theory
for the relevant matrix elements of $Q^S_6$ has been worked out.

As is evident from these equations, this useful decomposition does not work
in the truly quenched case where $N=0$.  In this case, Golterman and Pallante
propose defining a non-singlet operator $Q^{NS}_6$ through the equation:
\begin{equation}
Q_6 = Q^{QNS}_6 + \frac{1}{2} Q^{QS}_6.
\label{eq:g&p}
\end{equation}
where we have added the extra $Q$ to the superscript of the
general operator $Q^S_6$ defined in Eq.~\ref{eq:q6s} to emphasize
that it is being defined for the quenched ($N=0$) case.  At first
glance, Eq.~\ref{eq:g&p} might offer the possibility of
distinguishing two distinct contributions to $Q_6$: due to
$Q^{QNS}_6$ and $Q^{QS}_6$, based upon their different chiral
transformation properties under the graded symmetry.  In turn this
separation could -- perhaps -- be used to motivate an alternative
definition of the quenched approximation, in which the non-singlet
piece is dropped.  Unfortunately, the decomposition in
Eq.~\ref{eq:g&p} is entirely arbitrary.  The ``non-singlet"
operator, $Q_6^{QNS}$ is not protected by the graded symmetry from
mixing with the singlet operator.  It is therefore ambiguous to
separate out the ``singlet piece" of $Q_6$; any amount of
$Q_6^{QS}$ could be added to the definition of $Q_6^{QNS}$,
allowing the coefficient of 1/2 in Eq.~\ref{eq:g&p} to be replaced
by any arbitrary number.  In the representation theory language of
the previous section, $Q^{QS}_6$ is actually contained within the
representation to which $Q^{QNS}_6$ belongs: $1(n|n) \subset
\overline{\rm adj}(n|n)$. Thus, we cannot use Eq.~\ref{eq:g&p} to
identify a possibly preferred ``singlet'' part of the original
$Q_6$ operator.  It should be noted that $Q_6^{QS}$ is protected
by the graded symmetry from mixing with non-singlet operators; our
conclusion due to the preceding arguments is simply that, since it
is impossible to unambiguously define the singlet piece of $Q_6$,
any such approach is difficult to motivate physically.

For concreteness, we have focused on the specific operator $Q_6$.  However,
the transformation properties of the other three gluonic operators are
quite similar. The operator $Q_5$ transforms in an identical fashion as
does $Q_6$ and these will of course mix when the energy scale at
which they are defined is changed.  The operators $Q_3$ and $Q_4$ are
somewhat different since they transform only under $SU(n|n-N)_L$.  While
this increases the number of representations that can appear, the absence of
right-handed indices makes their quenched chiral perturbation theory less
singular.

We will now exploit these quenched chiral symmetry properties to
study the matrix elements of these operators in the chiral limit.

\subsection{Review of quenched ChPT}
\label{subsec:QChPT}

We now focus on the case of central interest in this work:
quenched QCD, as above concentrating on the operator $Q_6$.  Since
there are separate ChPT predictions for the matrix elements of the
operators, $Q_6$ and $Q_6^{QS}$ we will analyze them both in this
paper.  To leading order in QChPT these operators can be
represented by
\begin{eqnarray}\label{eq:penguin_qs}
Q_6^{QS} &=&
    \alpha_{q1}^{QS}\textrm{str}[\lambda_{6}\partial_{\mu}\Sigma\partial^{\mu}\Sigma^{\dag}]
    +2\,\alpha_{q2}^{QS} B_{0}\,
    \textrm{str}[\lambda_{6}(M \Sigma^\dagger + \Sigma M^\dagger)] + \textrm{h.c.}, \\
\label{eq:penguin_qns}
Q_6 &=&
   \frac{1}{2} \{\alpha_{q1}^{(8,1)}\textrm{str}[\lambda_{6}\partial_{\mu}\Sigma\partial^{\mu}\Sigma^{\dag}]
    +2\,\alpha_{q2}^{(8,1)} B_{0}\,
    \textrm{str}[\lambda_{6}(M \Sigma^\dagger + \Sigma M^\dagger)] + \textrm{h.c.}\} \nonumber \\
    & &+\alpha_{q}^{NS}\textrm{str}[\lambda_{6}\Sigma \bar A \Sigma^{\dag}]+ \textrm{h.c.}
\end{eqnarray}
where $(\lambda_6)_{ij}= \delta_{i3}\delta_{j2}$ while
\begin{eqnarray}
M         &=& (m_u,m_d,m_s,m_u,m_d,m_s)_{\rm diag}, \\
\bar A &=& (1,1,1,-1,-1,-1)_{\rm diag}, \\
B_{0}        &=& \frac{m^{2}_{\pi^{+}}}{m_{u}+m_{d}}= \frac{m^{2}_{K^{+}}}{m_{u}+m_{s}}
                  =\frac{m^{2}_{K^{0}}}{m_{d}+m_{s}}.
\end{eqnarray}
The matrix $\bar A$ is an element of the extended adjoint
representation $\overline{\rm adj}(n|n)$ and appears here because
the corresponding operator in the underlying quenched theory,
defined in Eq.~\ref{eq:g&p}, transforms in the same fashion.  The
meson field $\Sigma$ is defined by
\begin{equation}
\Sigma = \exp \left[\frac {2i\Phi}{f}\right]
\end{equation}
with
\begin{equation}
\Phi \equiv \left(
\begin{array}{cc}
  \phi & \chi^{\dag} \\
  \chi & \widetilde{\phi} \\
\end{array}%
\right).
\end{equation}
The quantity $f$ is the pseudoscalar decay constant in the chiral
limit (In the normalization used the physical value of the pion
decay constant is $f_\pi\simeq 130$ MeV) while the $3 \times 3$
matrices $\phi$, $\widetilde{\phi}$ and $\chi^\dag$ are
constructed from Goldstone fields which create and destroy
particles made from valence quarks and anti-quarks (bosons),
ghost-anti-ghost quarks (bosons) and quarks and anti-ghost quarks
(fermions) respectively.

We note that the ChPT representation for the operator $Q_6$ given
in Eq.~\ref{eq:penguin_qs} differs from the implications of the
original formula of Golterman and Pallante, Eq.~3.5 of
Ref.~\cite{Golterman:2001qj}.  In particular, the quenched
non-singlet operator $Q_6^{QNS}$ will be represented in chiral
perturbation theory by both singlet and non-singlet operators.  As
a result, the low energy constants $\alpha^{(8,1)}_{q1}$ and
$\alpha^{(8,1)}_{q2}$ which multiply the two quenched singlet
operators which appear in $Q_6$ need not agree with the
coefficients $\alpha^{QS}_{q1}$ and $\alpha^{QS}_{q2}$ which
appear in the singlet operator $Q^{QS}_6$.

In Eqs.~\ref{eq:penguin_qs} and \ref{eq:penguin_qns} and those below,
we combine the $2n \times 2n$ matrix $\Sigma$ together with the similar
matrices $\bar A$ and $M$, which appear in the quark-level theory,
to form the most general set of operators that are invariant under the
complete graded symmetry $S(n|n)_L \times SU(n|n)_R$ of the quenched
theory, to a given order in the Goldstone particle masses and momenta.
As matrices transforming in the product representation
$(n|n) \times \overline{(n|n)}$ they can be written as:
\begin{equation}
U = \left(
\begin{array}{cc}
  A & B \\
  C & D \\
\end{array}
\right),
\end{equation}
where the sub-matrices have the same dimension as the sub-matrices
of $\Phi$, above.  The invariance under the graded symmetry of the
quenched theory requires the presence of supertraces in the
operators of Eqs.~\ref{eq:penguin_qs} and \ref{eq:penguin_qns}
defined by
\begin{equation}\label{eq:def_str}
\textrm{str}(U)=\textrm{tr}(A)-\textrm{tr}(D).
\end{equation}

To tree-level in ChPT, a pseudoscalar meson mass is given by
\begin{equation}\label{11}
m_{ij}^2 = B_{0}(m_i + m_j),
\end{equation}
where $m_i$ and $m_j$ are the masses of the two quarks
that form the meson.  We define $m_{33}$ to be the tree-level
meson mass of two valence strange quarks, as in \cite{Golterman:2000fw}
\begin{equation}\label{12}
m^2_{33}=2m^2_K-m^2_\pi.
\end{equation}

The new non-singlet operator of Eq.~\ref{eq:penguin_qns} is nominally of
$O(p^0)$ in ChPT, but its tree-level contributions to physical matrix elements
vanish because $\bar A$ is proportional to the unit matrix in the
valence sector \cite{Golterman:2002us}.  At $O(p^2)$ this is not true and
the one-loop insertions of $Q_6^{QNS}$ make a contribution
of the same order as the tree-level insertions of $Q_6^{QS}$.  Of
course, there are additional non-singlet operators that can be constructed
from the matrix $\bar A$ which enter at $O(p^2)$ whose presence is
needed to compensate for the scale dependence of the one-loop insertions of
$Q_6^{QNS}$.  This introduces three more LEC's into the amplitudes
we must consider in this paper.  The effective Lagrangian to this order is
\cite{Golterman:2002us}:

\begin{equation}\label{13}
    {\cal L}^{(NLO)}_{NS}=  \sum_{i}
    c_i^{NS}{\cal O}^{NS}_{i},
\end{equation}

\noindent with

\begin{equation}\label{14}
\begin{array}{ll}
{\cal O}^{NS}_{1}= \textrm{str}[\lambda_{6} L_{\mu}\Sigma^{\dag} \bar A \Sigma L^{\mu}],\\
{\cal O}^{NS}_{3}= \textrm{str}[\lambda_{6} \{\Sigma^{\dag} \bar A \Sigma, L^{2}\}],\\
{\cal O}^{NS}_{4}= \textrm{str}[\lambda_{6} \{\Sigma^{\dag} \bar A \Sigma, S\}],\\
\end{array}
\end{equation}

\noindent and $ S=2B_{0}(M^{\dag}\Sigma + \Sigma^{\dag}M$),
$L_{\mu}=i \Sigma^{\dag}\partial_{\mu}\Sigma$.  In the following
we work with operators renormalized in the $\overline{MS}$ scheme,
absorbing the divergence into the coefficients of the effective
Lagrangian, which have the form

\begin{equation}\label{15}
    c_{i}^{NS}=c^{r,NS}_{i}+\frac{1}{16
    \pi^{2}f^{2}}\left[\frac{1}{d-4}+\frac{1}{2}(\gamma_{E}-\ln 4
    \pi)\right]2 \alpha_{q}^{NS} \eta_{i}.
\end{equation}

The finite coefficients, $c_i^{r,NS}$, are the renormalized low
energy constants of the theory, while the factors $\eta_i$ are
chosen to cancel the divergences of the one-loop insertions of the
tree-level operator.  In the quenched theory we find $\eta_1=0$,
$\eta_3=3$ and $\eta_4=-3$.  The scale dependence of the LEC's is
given to one-loop order by
\begin{equation}\label{16}
c^{r,NS}_i(\mu_2)=c^{r,NS}_i(\mu_1)+\frac{2\alpha_{q}^{NS}\eta_i}{(4\pi
f)^2}\ln{\frac{\mu_1}{\mu_2}},
\end{equation}
where $\mu_1$ and $\mu_2$ are two different values of
the chiral scale.  The result for physical amplitudes should be
independent of the scale and Eq.~\ref{16} is obtained by
requiring that all one-loop amplitudes for the $Q_6$ operator be
scale independent when the $O(p^2)$ NLO LEC's are included in the
calculation.

The relations between our conventions for the LEC's and those of
Golterman and Pallante \cite{Golterman:2001qj,Golterman:2002us} are given below.
The constant $\alpha_q^{NS}$ has been chosen to have a
normalization that agrees with that of $\alpha_q^{(8,8)}$ in
Ref.~\cite{Blum:2001xb}:
\bea\label{17} \alpha_{q,GP}^{NS} =
\frac{2}{f^2}\alpha_q^{NS}.\eea
In addition, we are working with the notation of
Ref.~\cite{Laiho:2003uy} for the NLO LEC's.  Although that work
dealt with the electro-weak penguins, both transform in the
$\bigl({\rm adj}(n|n-N)_L,{\rm adj}(n|n-N)_R\bigr)$ representation
of the partially quenched graded symmetry in the case $N \ne 0$
and we therefore keep that notation. The relationship between the
two is

\begin{eqnarray}\label{18}
\beta_{q1}^{NS}=(4\pi)^2 2c_3^{NS}, \nonumber \\
\beta_{q2}^{NS}=(4\pi)^2 2c_1^{NS},  \nonumber \\
\beta_{q3}^{NS}=(4\pi)^2 2c_4^{NS}.
\end{eqnarray}

\subsection{$Q_6$ amplitudes in quenched ChPT}

We now review the ingredients necessary to obtain the contribution
of both the operators $Q^{QS}_6$ and $Q_6$ to the $K\to\pi\pi$
amplitude to leading order, $O(p^2)$, in quenched ChPT.  The
chiral behavior of $Q_6^{QS}$ is the same as that of $Q_6$ in the
full theory:
\bea\label{19} \langle
\pi^{+}\pi^{-}|Q^{QS}_6|K^{0}\rangle & =&
\frac{4i\alpha_{q1}^{QS}}{f^{3}}(m^2_K-m^2_\pi),  \eea
where, as in the full QCD case, the needed LEC $\alpha_{q1}^{QS}$
can be extracted from the $K\to0$ and $K\to\pi$ matrix elements:
\begin{eqnarray}\label{20}
\langle 0|Q^{QS}_6|K^{0}\rangle &=&
\frac{4i\alpha_{q2}^{QS}}{f}(m^{2}_{K}-m^{2}_{\pi}),
\end{eqnarray}
\begin{eqnarray}\label{21}
\langle \pi^{+}|Q^{QS}_6|K^{+}\rangle &=&
\frac{4}{f^{2}}\alpha_{q1}^{QS}m_{K}m_{\pi} -
\frac{4}{f^{2}}\alpha_{q2}^{QS}m^{2}_{K}.
\end{eqnarray}

The $K\to\pi\pi$ matrix element of $Q_6$ includes the one-loop
contributions of $\alpha_q^{NS}$ and the $O(p^2)$ LEC's, $c_i^{NS}$.
In this case, $K\to\pi\pi$ is given by \cite{Golterman:2002us}:
\bea\label{22} \langle \pi^{+}\pi^{-}|Q_6|K^{0}\rangle &
=&
\frac{4i}{f^{3}}\left(\frac{1}{2}\alpha_{q1}^{(8,1)}-c_1^{NS}-2c_3^{NS}\right)
(m^2_K-m^2_\pi) \nonumber
\\ && +\frac{2i}{16\pi^2f^5}\alpha_q^{NS}\left[12(m_K^2-m_\pi^2)
\left(\ln\frac{m^2_\pi}{\mu^2}-1\right) \right. \nonumber \\ &&
 +\left(\frac{m_K^6}{m_\pi^4}-2\frac{m_K^4}{m_\pi^2}+2m^2_K
\right)\ln\frac{m^2_K}{m^2_\pi} \nonumber \\ &&
+\left(\frac{m_K^6}{m^4_\pi}-6\frac{m^4_K}{m^2_\pi}+10m^2_K-4m^2_\pi
\right)\ln\frac{m^2_{33}}{m^2_\pi} \nonumber \\ &&
+2m^2_K\biggl(F(m^2_\pi,m^2_\pi,-m^2_K)-2i\pi
\theta(m^2_K-4m^2_\pi) \biggr. \nonumber \\ &&
\left.\times\sqrt{1-\frac{4m^2_\pi}{m^2_K}}+\frac{\pi}{3}\sqrt{3}
\right)+\left(\frac{m^4_K}{m^2_\pi}-2m^2_K \right) \nonumber \\ &&
\biggl.
\times\left(2F(m^2_\pi,m^2_K,-m^2_\pi)+F(m^2_K,m^2_{33},-m^2_\pi)
\right)\nonumber \\ && +6m^2_K\left(1-\frac{m^2_K}{m^2_\pi}\right)
\biggr], \eea
where the function $F$ is given in Appendix A.  Note the scale
dependence in the logarithmic term proportional to
$\alpha_q^{NS}$. This scale dependence cancels that of $c_3^{NS}$,
as can be seen from Eq.~\ref{16}.  The $Q_6$ amplitudes for
$K\to0$ and $K\to\pi$ are \cite{Golterman:2002us}:
\begin{eqnarray}\label{23}
\langle 0|Q_6|K^{0}\rangle &=&
\frac{4i}{f}\left(\frac{1}{2}\alpha_{q2}^{(8,1)}+2c_4^{NS}\right)(m^{2}_{K}-m^{2}_{\pi})
\nonumber \\ &&
+\frac{8i}{f^3}\alpha_q^{NS}\frac{1}{16\pi^2}\left[m^2_K\ln\frac{m^2_K}{\mu^2}
-2m^2_\pi\ln\frac{m^2_\pi}{\mu^2}+m^2_{33}\ln\frac{m^2_{33}}{\mu^2}
\right. \nonumber \\ && \biggl. -3(m^2_K-m^2_\pi) \biggr], \\
\label{24}
\langle \pi^{+}|Q_6|K^{+}\rangle &=&
\frac{4m_M^2}{f^{2}}\left(\frac{1}{2}\alpha_{q1}^{(8,1)} -
\frac{1}{2}\alpha_{q2}^{(8,1)}-c_1^{NS}-2c_3^{NS}-2c_4^{NS}\right),
\end{eqnarray}
where in the expression for $K\to\pi$ we have set the quark masses
to be degenerate $(m^2_K=m^2_\pi=m^2_M)$, as in the numerical
simulation \cite{Blum:2001xb}. Note, as discussed above, the
singlet LEC's $\alpha_{q1}^{(8,1)}$ and $\alpha_{q2}^{(8,1)}$
entering the equations for $Q_6$ (Eqs.~\ref{22}, \ref{23} and
\ref{24}) need not be the same as the corresponding LEC's,
$\alpha_{q1}^{QS}$ and $\alpha_{q2}^{QS}$appearing in the
corresponding expressions for $Q^{QS}_6$ (Eqs.~\ref{19}, \ref{20}
and \ref{21}).

Notice that the same linear combinations of LEC's appear in the
above three equations, $\alpha_{q1}^{(8,1)}/2-c_1^{NS}-2c_3^{NS}$
and $\alpha_{q2}^{(8,1)}/2+2c_4^{NS}$.  If the value of
$\alpha_q^{NS}$ is small, then the procedure for obtaining
$K\to\pi\pi$ in which one neglects $\alpha_q^{NS}$ reduces to that of
the full theory. In this case, the previously mentioned combinations of LEC's
replace $\alpha_1^{(8,1)}$ and $\alpha_2^{(8,1)}$, respectively.  The
strategy that was employed in Refs.~\cite{Noaki:2001un} and \cite{Blum:2001xb}
implicitly made this approximation.

It has recently been suggested, however, that the value of
$\alpha_q^{NS}$ is large compared to the other LEC's in the
amplitudes and cannot be neglected \cite{Golterman:2003uj}.  A large
$N_{c}$ expansion was used to obtain the result (in our
conventions),
\bea\label{25} \alpha^{NS}_q = -\frac{1}{4} f^4 B_0^2. \eea
When quenched lattice values for these constants are
substituted into this equation, one finds that the resulting value
of $\alpha_q^{NS}$ is so large that it cannot be ignored.  The
non-linearities implied by the presence of $\alpha_q^{NS}$ in
Eq.~\ref{23} need to be included when this expression is used
to extract the LEC's from the $K\to 0$ lattice data and the explicit
form of the $\alpha_q^{NS}$-dependent term in Eq.~\ref{22} needs to be
taken into account in a quenched prediction for $K\to\pi\pi$.
Of course, with such large quenching artifacts, such a quenched
prediction will be of limited value.

It is therefore crucial to have a means for determining
$\alpha_q^{NS}$ directly from a lattice calculation.  Golterman
and Pallante have provided a method for doing just that.  However,
we have found that their method suffers from ambiguities which are
power divergent in the lattice spacing when employed in the case
of domain wall fermions, making accurate extraction of
$\alpha_q^{NS}$ from the lattice data rather difficult. In
particular, with a finite $L_s$ (the separation between the two
physical, four-dimensional walls in the fifth dimension), this
power divergent contribution, although suppressed by a factor of
order of the residual chiral symmetry breaking, $O(m_{\rm res})$,
could be large.  We discuss this difficulty in the following
sections and present an alternative method for obtaining
$\alpha_q^{NS}$ which avoids this problem.

\section{Lattice determination of $\alpha^{NS}_q$ from
$\widetilde{K}\to0$}
\label{sec:alpha_NS_K_vac_thry}

In this section we review the method proposed by Golterman and
Pallante to obtain $\alpha_q^{NS}$ on the lattice
\cite{Golterman:2002us}, together with the (CPS) symmetry
arguments needed to understand the form of the power divergences
in the matrix elements of four-quark operators.  As will be
explained in the following, Golterman and Pallante introduced a
new operator, $\widetilde{Q}^{QNS}_6$, and extracted
$\alpha_q^{NS}$ from matrix elements of this operator which
included ghost particles in the external states.  By extending the
standard CPS symmetry arguments to include quark-ghost
transformations, we show that this method suffers from
contamination from terms which are power divergent in the lattice
spacing.  Nevertheless, we present numerical results from this
approach.  While these results are in rough agreement with the
large $N_c$ estimates \cite{Golterman:2003uj}, due to the
ambiguities associated with the power divergence the results must
be considered inconclusive; in Section VI we suggest an
alternative approach which does not suffer from this problem.

\subsection{Review of Golterman and Pallante's method}

In Eq.~\ref{23} it is difficult to numerically disentangle the
logarithmic $\alpha_q^{NS}$ term from the linear term and possible
higher order effects.  It was for this reason that a new matrix element
was suggested in Ref.~\cite{Golterman:2002us} to which $\alpha_q^{NS}$
contributes at $O(p^0)$, so that it may be obtained more readily in
the chiral limit. This can be done if one considers a matrix element
where the ghost quarks can appear on external lines.  In order to obtain
such a matrix element, one must perform an $SU(3|3)_L$ flavor
rotation of the operator $Q_6^{QNS}$ into

\bea \label{26}
\widetilde{Q}^{QNS}_6
      &=& -\frac{1}{2}\overline{s}\gamma^\mu(1-\gamma^5)\widetilde{d}
          \left(\sum_{q}\overline{q}\gamma^\mu(1+\gamma^5)q -
             \sum_{\widetilde{q}} \overline{\widetilde{q}}\gamma^\mu(1+\gamma^5)\widetilde{q}\right)
                                                                     \nonumber \\
      &=& -\frac{1}{2}\overline{{\cal Q}}\,\widetilde{\lambda}_6\gamma^\mu(1-\gamma^5){\cal Q}\;
                        \overline{{\cal Q}}\,\bar A\gamma^\mu(1+\gamma^5){\cal Q},
\eea
where now the $\widetilde{d}$ is a ghost quark field.  The matrix
$\widetilde{\lambda}_6$ is given by $(\tilde{\lambda}_6)_{ij} =
\delta_{i 3}\delta_{j5}$, a quenched chiral transform of the
matrix $\lambda_6$ defined earlier.  To leading order in ChPT,
this operator is

\bea \label{26.5}  \widetilde{Q}^{QNS}_6
      &=&\alpha_{q}^{NS}\textrm{str}[\widetilde{\lambda}_{6}\Sigma \bar A \Sigma^{\dag}]+
      \textrm{h.c.} \eea

\noindent  Note that some care must be taken in order to maintain
consistency in the sign conventions between the chiral and quark
level operators.  Our conventions are presented in detail in
Appendix B. Since the above operator is in the same irreducible
representation as $Q^{QNS}_6$, it is parameterized by the same low
energy constants. Considering the matrix element
$\widetilde{K}\to0$, we have, to leading order
\begin{eqnarray}\label{27}
\langle 0|\widetilde{Q}^{QNS}_6|\widetilde{K}^{0}\rangle
&=& \frac{4i}{f}\alpha_{q}^{NS}. \eea

Although this method isolates $\alpha_q^{NS}$ at leading order,
the NLO contribution has a power divergent coefficient, making the
numerical extraction problematic.  We discuss the way in which the
mixing of four-quark operators with power divergent lower
dimensional operators can be constrained by the symmetries of the
theory in the following subsection.

\subsection{Power divergences and CPS symmetry}
\label{subsec:quad_dvrgt}

In general, the $\Delta I =1/2$ matrix elements of four-quark
operators have a power divergent part, due to mixing with lower
dimensional operators.  This power divergence will involve the
quark bilinears, $\overline{s}d$ and $\overline{s}\gamma_5d$,
times a momentum independent coefficient \cite{Blum:2001xb}.
One can define the following quark bilinear operator
\cite{Bernard:1985wf},
\begin{equation}\label{28}
\Theta^{(3,\overline{3})}\equiv \overline{s}(1-\gamma_5)d,
\end{equation}
which is equal to
$\alpha^{(3,\overline{3})}\textrm{Tr}(\lambda_6\Sigma)$ to lowest
order in ChPT, where in our conventions,
$\alpha^{(3,\overline{3})}=-\frac{f^2}{2}B_0$.

We briefly review the use of CPS symmetry \cite{Bernard:1987pr} to
determine the form of the power divergences that enter our
computations due to mixing with the lower dimensional operator of
Eq.~\ref{28}.  Here $C$ and $P$ are the usual charge conjugation
and parity inversion symmetries, while $S$ is the symmetry under
exchange of the $s$ and $d$ quarks, which is exact when the quark
masses are equal.

The parity even part of the above operator, $\overline{s}d$, has a
CPS of $+1$, while the parity odd part, $\overline{s}\gamma_5d$,
has a CPS of $-1$.  The operator, $Q^{QNS}_6$, has a CPS
of $+1$ and the matrix element $\langle 0|Q^{QNS}_6|K\rangle$
is a parity odd transition.  Therefore, the power divergence of
this matrix element must be proportional to the matrix element of
a lower dimensional operator which is also parity odd and has
CPS=$+1$. We see that the only operator with these symmetries that
can be constructed is $\overline{s}\gamma_5d$ multiplied by
$m_s-m_d$. Thus, the power divergent part of $\langle
0|Q^{QNS}_6|K\rangle$ is proportional to $m_s-m_d$.

For the parity even transition, $\langle \pi|Q^{QNS}_6|K\rangle$,
the mixing must be with a parity even lower dimensional operator
with CPS$=+1$. In this case the only such operators are
$\overline{s}d$ and $(m_s+m_d)\overline{s}d$.  The former is ruled
out by chiral symmetry.  As domain wall fermions at finite $L_s$
break chiral symmetry such a mixing can occur, albeit greatly
suppressed due to the domain wall fermion mechanism. Thus, the
dominant part of the power divergence of the $K\to\pi$ matrix
element is proportional to $m_s+m_d$. Since this is a statement at
the operator level, it is true to all orders in the chiral
expansion, an important result when dealing with the numerical
data of a lattice calculation.

We now wish to consider matrix elements of the operator
$\widetilde{Q}^{QNS}_6$, since this will give us
$\alpha_q^{NS}$ at $O(p^0)$ in the chiral expansion.  In order to
proceed, we construct a new graded symmetry,
CP$\widetilde{\textrm{S}}$, where now $\widetilde{\textrm{S}}$ not
only exchanges $s$ and $d$ quarks, but also exchanges all valence
quarks with their corresponding ghosts, {\it i.e.} under
$\widetilde{\textrm{S}}$ the six ``quark'' fields transform as
\begin{equation}
(u,d,s,\tilde{u},\tilde{d},\tilde{s})
                  \to (\tilde{u},\tilde{s},\tilde{d},u,s,d).
\end{equation}
In order to implement
this CP$\widetilde{\textrm{S}}$ symmetry, we need to understand
how $C$ and $P$ act on ghost quark states.

Under $C$, the valence quarks have the following familiar
transformation properties
\bea
\label{29}            q  &\rightarrow& C\overline{q}^T, \\
\label{30}  \overline{q} &\rightarrow& -q^TC^{-1}.
\eea
Thus, the quark bilinears, $\overline{s}d$ and
$\overline{s}\gamma_\mu d$ transform as,
\bea
\label{31} \overline{s}d            &\rightarrow& \overline{d}s, \\
\label{32} \overline{s}\gamma_\mu d &\rightarrow&
                                      -\overline{d}\gamma_\mu s,
\eea
where we have used the identities, $C^\dag=C^{-1}$ and
$C^{-1}\gamma_\mu C=-\gamma_\mu^{T}$ and the fact that the quark
fields anti-commute.

In the case of ghost quarks we can define $C$ analogously to the
case of valence quarks, such that
\bea\label{33}  \widetilde{q}\rightarrow
C\overline{\widetilde{q}}^T. \eea
However, the anti-ghost does not transform independently
from the ghost, because it obeys bose statistics.  Taking the
transpose conjugate of both sides yields
\bea\label{34}  \overline{\widetilde{q}}\rightarrow
\widetilde{q}^TC^{-1}, \eea
which differs in sign from Eq.~\ref{30}.  Using the
transformation rules in Eqs.~\ref{33} and \ref{34} and the fact
that ghost quark fields commute, we see that
\bea\label{34.5} \overline{\widetilde{s}}\gamma_\mu \widetilde{d}
\rightarrow -\overline{\widetilde{d}}\gamma_\mu \widetilde{s},
\eea
where the sign change demonstrates that this
transformation for ghost quarks does indeed act as a charge
conjugation.  For bilinears that are made of a valence quark and
an antighost quark the fact that the fields commute will cause a sign
change compared to Eqs.~\ref{31} and \ref{32}. For example,
\bea\label{35} \overline{s}\gamma_\mu \widetilde{d} \rightarrow
\overline{\widetilde{d}}\gamma_\mu s. \eea

Under $P$, the quark fields transform as
\bea
\label{36}           q  &\rightarrow& P q, \\
\label{37} \overline{q} &\rightarrow& \overline{q}P,
\eea
where $P^\dag P=1$.  It is simple to recognize that the same
transformation under parity is also valid for the ghost fields.

We are now ready to examine the CP$\widetilde{\textrm{S}}$
transformation properties of the relevant operators.  The
bilinear, $\overline{s}\widetilde{d}$, has a
CP$\widetilde{\textrm{S}}$ of $-1$, while
$\overline{s}\gamma_5\widetilde{d}$ has a
CP$\widetilde{\textrm{S}}$ of $+1$.  The four quark operator,
$\widetilde{Q}^{QNS}_6$, has a CP$\widetilde{\textrm{S}}$ of $+1$.
Thus, the matrix element, $\langle
0|\widetilde{Q}^{QNS}_6|\widetilde{K}\rangle$, which is a parity
odd transition, must have a power divergence given by the matrix
element of a lower dimensional operator which is both parity odd
and has CP$\widetilde{\textrm{S}}$=$+1$.  We see that the bilinear
operator $\overline{s}\gamma_5\widetilde{d}$, which already has
the same CP$\widetilde{\textrm{S}}$ as $\widetilde{Q}^{QNS}_6$,
can only have the coefficient $m_s+m_d$.  This is opposite to the
situation for the $K\to0$ amplitude, where the CPS of $Q^{QNS}_6$
and $\overline{s}\gamma_5 d$ was opposite, implying a power
divergence proportional to $m_s-m_d$. The reason for the
difference is the exchange of valence and ghost quarks in the
right-handed part of Eq.~\ref{26} under $\widetilde{\textrm{S}}$,
yielding an extra relative minus sign.

Note, this same conclusion could also be reached by starting with
the dimension-3 operator which represents the quadratic divergence
present in the usual operator $Q_6$:
\begin{equation}
\overline{q}\lambda_6 M (1+\gamma^5)q
  + \overline{q}M^\dagger \lambda_6 (1-\gamma^5)q
   = (m_d+m_s)\overline{s}d + (m_d-m_s)\overline{s}\gamma^5d.
\label{eq:quad_dvrgt}
\end{equation}
Note, the form of the left-hand-side of Eq.~\ref{eq:quad_dvrgt} is
determined by the chiral symmetry properties of the matrix
$\lambda_6$ (an element of $(8,1)$) and the quark mass matrix $M$
(an element of $(3,\overline{3})$).  The relative sign of the two
terms on the left-hand-side of this equation is determined by CPS
symmetry.  The corresponding dimension-3 operator that mixes with
$\widetilde{Q}^{QNS}_6$ must have the same form except for the
addition of the matrix $\bar A$ and a quenched chiral rotation of
$\lambda_6$ to $\widetilde{\lambda_6}$ to agree with the quenched
chiral symmetry properties of $\widetilde{Q}^{QNS}_6$:
\begin{equation}
\overline{\cal Q}\,\widetilde{\lambda}_6 M \bar A (1+\gamma^5){\cal Q}
  + \overline{\cal Q}\, \bar A  M^\dagger \widetilde{\lambda}_6 (1-\gamma^5){\cal Q}
   = (m_s-m_d)\overline{s}\widetilde{d}
   - (m_s+m_d)\overline{s}\gamma^5\widetilde{d}.
\label{eq:quad_dvrgt_q}
\end{equation}

This argument demonstrates that $\widetilde{K}\to0$ has a power
divergence proportional to $m_s+m_d$.  When the fermion
discretization does not preserve exact chiral symmetry (such as
domain wall fermions with finite $L_s$), there will also be mixing
with the operator $\overline{s}\gamma^5 d$.  For domain wall fermions this
is suppressed by a power of $m_{res}$.  Such power divergences
lead to large uncertainties when one tries to obtain the chiral
limit of a matrix element without exact chiral symmetry on the
lattice. However, these symmetry arguments suggest a way around
this difficulty and we provide an alternative method for
obtaining $\alpha_q^{NS}$ in Section~\ref{sec:alpha_NS_K_pi_thry}
which makes use of the following observation: the matrix element
$\langle \widetilde{\pi}|\widetilde{Q}^{QNS}_6|K\rangle$ is parity
even and since the parity even bilinear, $\overline{s}\widetilde{d}$,
has CP$\widetilde{\textrm{S}}=-1$, the divergence of
$K\to\widetilde{\pi}$ must be proportional to $m_s-m_d$.  For
degenerate quark masses, the power divergence vanishes completely.
Again, this is a statement at the operator level and holds to
all orders in the chiral expansion.

\subsection{Lattice contractions for Golterman and Pallante's method}
\label{subsec:G&P_contract}

Continuing the derivation of Ref.~\cite{Golterman:2002us}, we carry out the
Wick contractions for $\langle
0|\widetilde{Q}^{QNS}_{penguin}|\widetilde{K}^{0}\rangle$, where
we specialize to the case, $Q^{QNS}_{penguin}=Q^{QNS}_6$,
\bea \label{38} \langle
0|\widetilde{Q}^{QNS}_{6}|\widetilde{K}^{0}\rangle &=&
-i\sum_{q\in\{u,d,s\}} [I_M(V,A,q)-I_M(A,V,q)] \\
 && +\frac{1}{2}\left[iI_M'(V,A,d)-iI_M'(A,V,d)+iI_M'(V,A,s)-iI_M'(A,V,s)\right],
    \nonumber
\eea
with
\bea\label{39} I_M(j,k,q) & \equiv &
\textrm{Tr}_c\{\textrm{Tr}_s[\Gamma_j S_d(x_{op},x_0) S_s(x_{op},x_0)^\dag\gamma_5]\nonumber
\\ && \times \textrm{Tr}_s[\Gamma_k S_q(x_{op},x_{op})]\}, \\
\label{40} I_M'(j,k,q') & \equiv &
\textrm{Tr}_s\{\textrm{Tr}_c[\Gamma_j S_d(x_{op},x_0) S_s(x_{op},x_0)^\dag\gamma_5]\nonumber
\\ && \times \textrm{Tr}_c[\Gamma_k S_{q'}(x_{op},x_{op})]\}.
\eea
In these contractions, $\Gamma_V=\gamma_\mu$ and
$\Gamma_A=\gamma_\mu\gamma_5$ and the traces are over spin or
color. The quantity $S_q(x,y)$ is the Dirac propagator connecting
positions $x$ and $y$ for a quark of type $q=(u,d,s)$. The
position $x_0$ locates the source for the $\widetilde{K}^0$ meson
while $x_{op}$ is the position of the operator
$\widetilde{Q}^{QNS}_6$. The derivation of Eq.~\ref{38} makes use
of the fact that ghost and valence propagators are equal flavor by
flavor (when properly ordered; see Appendix B).  As discussed in
Ref.~\cite{Golterman:2002us}, this is important because it allows
us to obtain $\alpha_q^{NS}$ from contractions that have already
been computed for $\langle 0|Q^{QCD}_{6}|K^{0}\rangle$.

Golterman and Pallante make the additional observation that the
linear combination of contractions for $\langle
0|Q^{QNS}_{6}|K^{0}\rangle$ is the same as Eq.~\ref{38}, but with
opposite sign for all but the $I_M'(j,k,d)$ terms.  However, a
careful treatment of the conventions for external ghost states
shows that the $\widetilde{K}\to 0$ matrix element has the
opposite sign compared to that given in Golterman and Pallante
\cite{Golterman:2003uj}, so that the contractions for $\langle
0|Q^{QNS}_{6}|K^{0}\rangle$ are the same as Eq.~\ref{38}, but with
opposite sign for \emph{only} the $I_M'(j,k,d)$ terms.  Thus,
since $K\to0$ does not contribute at $O(p^0)$, subtracting $K\to
0$ from $\widetilde{K}\to 0$ yields
\bea\label{41} \langle
0|\widetilde{Q}^{QNS}_{6}|\widetilde{K}^{0}\rangle - \langle
0|Q^{QNS}_{6}|K^{0}\rangle &=& i
[I_M'(V,A,d)-I_M'(A,V,d)]\nonumber \\ &&
=\frac{4i}{f}\alpha_{q}^{NS} + O(p^2). \eea
Again, we point out that the $O(p^2)$ terms are
multiplied by a quadratic divergence in the lattice spacing.

Using the leading order ChPT result,
\begin{equation}\label{42}
\langle 0|\Theta^{(3,\overline{3})}|K^{0}\rangle =
\frac{2i}{f}\alpha^{(3,\overline{3})}=- \langle 0 | \bar s\gamma_5
d | K^0 \rangle,
\end{equation}

\noindent we obtain the following ratio:

\bea\label{43} \frac{\langle
0|\tilde{Q}^{QNS}_6|\tilde{K}^0\rangle - \langle
0|Q^{QNS}_6|K^0\rangle}{\langle 0|\overline{s}\gamma_5
d|K^0\rangle} &=&
-2\frac{\alpha^{NS}_q}{\alpha^{(3,\overline{3})}} +
\frac{\textrm{const.}}{a^2}m_d+O(p^2).\nonumber \\ && \eea
Since we are dividing by precisely the matrix element that,
according to the operator level discussion of the previous
section, multiplies the power divergence, Eq.~\ref{43} explicitly
gives the form of the power divergence, with the neglected higher
order terms in the chiral expansion being free of these
divergences. Again, $\alpha_q^{NS}$ can be obtained in the chiral
limit. The power divergence of Eq.~\ref{43} is proportional to
$m_d$, since it is the sum of two terms, one proportional to
$m_s+m_d$ and the other one proportional to $m_s-m_d$, with the
$m_s$ term cancelling between them.  The only divergent term in
this ratio depends linearly on $m_d$ as shown in Eq.~\ref{43}.
However, after this term proportional to $m_d$ has been removed by
extrapolation to $m_d \to 0$ there remains an $O(m_{res}/a^2)$
contribution for domain wall fermions, due to the power divergence
and the lack of exact chiral symmetry for finite
$L_s$~\cite{Blum:2001xb}.  For our simulations, this effect could
be as large as the constant we are trying to extract, as we show
in the next subsection.

\section{Numerical results for $\alpha^{NS}_q$ from $\widetilde{K}\to0$}
\label{sec:alpha_NS_K_vac_num}

We now examine the ratio of matrix elements of Eq.~\ref{43}, which
can be obtained from the contractions given in Eq.~\ref{38} and
the value for $\langle 0|\overline{s}\gamma_5d|K\rangle$.  We
briefly review the details of the ensemble used in
Ref.~\cite{Blum:2001xb} to generate the needed contractions.

The quenched configurations were generated with the Wilson gauge
action at a coupling of $\beta=6.0$ with lattice volume
$16^3\times32$.  The ensemble is comprised of 400 configurations,
separated by 10,000 sweeps, with each sweep consisting of a simple
two-subgroup heat-bath update of each link (Cabibbo-Marinari with
Kennedy-Pendleton accept-reject step). The lattice cut-off was
$a^{-1}=1.922(40)$ GeV set by the $\rho$ mass. The domain wall
fermion fifth dimension was $L_s=16$ sites with a domain wall
height $M_5=1.8$. The resulting residual quark mass was 0.00124(5)
in lattice units. For comparison, the value of $m_f$ corresponding
to a pseudo-scalar state made of degenerate quarks with mass equal
to the physical kaon at $\beta=6.0$ is approximately 0.0185.  The
light quark masses in units of the lattice spacing were taken to
be $m_f=$ 0.01, 0.02, 0.03, 0.04 and 0.05.

The data for the ratio in Eq.~\ref{43} are given in
Table~\ref{tabone}, and are plotted in Fig.~\ref{fig6} as a
function of $m_d$. We see from Eq.~\ref{43} that the intercept of
this graph allows one to obtain $\alpha_q^{NS}$, modulo the
effects of the residual chiral symmetry breaking. The two lines in
Figure \ref{fig6} represent linear fits to the data, the top one
plotted as a function of $m_d$, while the bottom is the same data
plotted against $m_d+m_{res}$. While in the presence of residual
chiral symmetry breaking effects neither of these two methods is
known to be correct, comparing these two approaches allows us to
estimate the order of magnitude of the $O(m_{\rm res})/a^2$
ambiguity.  As can be seen the difference is of the order of 30\%,
suggesting a potentially large uncontrolled error due to the power
divergence.

The result of an uncorrelated fit to the form of
$\eta_0+\eta_1(m_d+m_{res})$ was: $\eta_0=-5.94(14)\times
10^{-3}$, $\eta_1=-2.0781(33)$, with a $\chi^2/\textrm{dof}$ of
$1.67(18)$. The value of $\alpha_q^{NS}$ obtained from this fit
was $-1.20(3)\times10^{-5}$, where the error is statistical only.
If one fits the data in Table~\ref{tabone} to a linear function of
$m_d$ (instead of $m_d+m_{res}$), then the $\alpha_q^{NS}$
obtained is $-1.72(3)\times10^{-5}$.  For comparison, the large
$N_{c}$ approximation, Eq.~\ref{25}, gives a value of
$\alpha_q^{NS}=-1.63(48)\times10^{-5}$ where, again, the error is
statistical only.  Here we have used the quenched lattice values
for $f$ and $B_0$ reported in the previous RBC works,
\cite{Blum:2001xb,Blum:2000kn}, obtained on lattices with the same
size and coupling as those used here to determine $\alpha_q^{NS}$.
In Ref.~\cite{Blum:2000kn} it was found that $f=0.0713(53)$ and
$B_0=1.59(3)$ in the same lattice units, for an earlier set of 85
configurations.  We see that there is rough agreement between the
large $N_c$ value and the range of values given directly by the
lattice $\widetilde{K}\to0$ matrix element in the chiral limit.
Since an $\alpha_q^{NS}$ of this size would cause significant
changes in the $K\to\pi\pi$ amplitude for $Q_6$, as mentioned
previously, it is crucial to remove the large systematic error due
to the residual power divergence.

\section{Lattice determination of $\alpha^{NS}_q$ from $K\to\widetilde{\pi}$}
\label{sec:alpha_NS_K_pi_thry}

As we proved in Sec.~\ref{subsec:quad_dvrgt}, the matrix element, $\langle
\widetilde{\pi}|\widetilde{Q}^{QNS}_6|K\rangle$, does not have any
power divergences when the quark masses are degenerate.  To NLO in
ChPT, it is given by
\bea\label{43.5} \langle
\tilde{\pi}^+|\tilde{Q}^{QNS}_6|K^+\rangle &=&
-\frac{4}{f^2}\alpha^{NS}_q + O(p^2). \eea
Again, we have an amplitude where $\alpha_q^{NS}$ can be obtained
in the chiral limit, this time without any power divergent
ambiguities. As discussed in the next section on numerical fits,
the NLO logarithmic term vanishes for this matrix element in the
degenerate mass case (See Appendix C for details of this
calculation). The combination of contractions needed for the new
amplitude (for degenerate quark masses) is
\bea\label{44} \langle \tilde{\pi}^+|\tilde{Q}^{QNS}_6|K^+\rangle
&=& -\frac{1}{2}[L^8_M(V)-L^8_M(A)] \nonumber
\\ && +\sum_{q\in(u,d,s)}[L^I_M(V,q)-L^I_M(A,q)], \eea
with
\bea\label{45} L_M^8(j) & \equiv &
\textrm{Tr}_s\{\textrm{Tr}_c[\Gamma_j S_d(x_{op},x_1) S_u(x_{op},x_1)^\dag\gamma_5]\nonumber
\\ && \times \textrm{Tr}_c[\Gamma_j S_u(x_{op},x_{0}) S_s(x_{op},x_0)^\dag\gamma_5]\}, \\
\label{46} L_M^I(j,q) & \equiv &
\textrm{Tr}_c\{\textrm{Tr}_s[\Gamma_j S_d(x_{op},x_1)\gamma_5 S_u(x_{1},x_0)S_s(x_{op},x_0)^\dag\gamma_5]\nonumber
\\ && \times \textrm{Tr}_s[\Gamma_j S_q(x_{op},x_{op})]\}. \eea
Here we use the same notation as in Eqs.~\ref{39} and \ref{40} and
again exploit the fact that the ghost and valence propagators are
equal flavor by flavor when properly ordered
\cite{Golterman:2002us}.

\section{Numerical results for $\alpha_q^{NS}$ from
$K\to\widetilde{\pi}$}
\label{sec:alpha_NS_K_pi_num}

The results for the
$\langle\tilde{\pi}|\tilde{Q}^{QNS}_6|K\rangle$ matrix element are
given in Table.~\ref{tabtwo} and plotted in Fig.~\ref{fig7}. The
chiral limit was obtained by a simple linear extrapolation of the
form $d_0 +d_1(m_f+m_{res})$, with $d_0=9.80(76)\times 10^{-3}$,
$d_1=0.624(17)$ and a $\chi^2/\textrm{dof}=0.10(11)$. This yields
a value of $\alpha_q^{NS}=-1.24(10)\times 10^{-5} $. In this
result we neglect the small error in $f^2$.  It turns out that the
linear fit is exact to NLO in ChPT, as a direct one-loop
calculation shows that in this degenerate mass case the
logarithmic term vanishes.  This calculation is discussed in
Appendix C.

A comparison between the two methods for obtaining
$\alpha_q^{NS}$, from $K\to\widetilde{\pi}$ and from
$\widetilde{K}\to0$, is shown in Figure \ref{fig8}, where
$f^2/(2\alpha^{(3,\overline{3})}) \langle
\tilde{\pi}^+|\tilde{Q}^{QNS}_6|K^+\rangle$ is plotted over the
$\widetilde{K}\to0$ results.  With this normalization the
$K\to\widetilde{\pi}$ matrix element given in Eq~\ref{43.5} has
the same analytical value for the chiral limit (as determined in
ChPT) as the $\widetilde{K}\to0$ formula given in Eqs~\ref{43}.
The $K\to\widetilde{\pi}$ result does not have a divergent part
and the smallness of this amplitude in the numerical data compared
to the $\widetilde{K}\to0$ amplitude reflects this fact. The
chiral limit agrees within statistical errors for both methods,
which is better than expected, given the systematic errors
associated with the $\widetilde{K}\to0$ amplitude.

The NLO contribution to the matrix element $K\to\widetilde{\pi}$
was computed in order to control the systematic error associated
with the extrapolation to the chiral limit.  The linear fit is
exact to NLO since the log term vanishes, and as Fig.~\ref{fig7}
illustrates, this fit is quite good.  We therefore conclude that
the error in our value of $\alpha_q^{NS}$ in lattice units is
dominated by the statistical uncertainty, which is around $10\%$.
As mentioned previously, these results are in rough agreement with
the large $N_c$ estimate of Ref.~\cite{Golterman:2003uj}, and so,
as these authors pointed out, the term proportional to
$\alpha_q^{NS}$ will have a large numerical effect on the matrix
elements of $Q_6$.


\section{Conclusion and Outlook}
\label{sec:conclusion}

The application of the quenched approximation to weak decay matrix
elements is made more challenging by the combination of 1) the
perturbative calculations (in which QCD vacuum polarization
effects are important) needed to connect the high energy scale
typical of the $W$, $Z$ and top masses with the low energy scale
of the actual decay and 2) the non-perturbative QCD calculations
(in which the quenched approximation might be employed) needed to
evaluate the relevant low energy matrix elements.  We have
demonstrated in some detail how the quenched approximation can be
applied to the latter without altering the calculations which
underlie the former.  The resulting approach is very close to that
employed in the two large-scale quenched calculations reported in
Refs.~\cite{Blum:2001xb, Noaki:2001un}.

We then reviewed the quenched chiral perturbation theory results
for the strong penguin amplitudes relevant for $\epe$ as presented
by Golterman and Pallante \cite{Golterman:2001qj,Golterman:2002us}.
They have shown that a new LEC, $\alpha_q^{NS}$, contributes to
quenched amplitudes of the strong penguin operators (specifically, to
$Q_6$) and they have proposed a lattice method to obtain this new constant.

In Ref.~\cite{Golterman:2003uj}, $\alpha_q^{NS}$ was calculated
using the large $N_c$ approximation and the value was found to
have a large effect on the $K\to\pi\pi$ matrix element of $Q_6$.
Such an effect could significantly alter the value of $\epe$
reported in Ref.~\cite{Blum:2001xb}. Therefore, we considered it
essential that $\alpha_q^{NS}$ be computed directly on the
lattice. The method proposed by \cite{Golterman:2002us} to use the
matrix element of $\widetilde{K}\to0$ to obtain $\alpha_q^{NS}$
from the lattice was implemented in this paper, but it was shown
that this method suffers from ambiguities due to power divergent
contributions when evaluated using the domain wall fermion
formalism.  We have shown that the amplitude,
$K\to\widetilde{\pi}$, provides an alternative method to obtain
$\alpha_q^{NS}$ from the lattice but without power divergent
contributions.  We implemented this method and obtained a value of
$\alpha_q^{NS}$ which is indeed large enough to have an important
effect on the $K\to\pi\pi$ matrix element for $Q_6$, and,
therefore, also on the quenched determination of $\epe$.

We conclude that we cannot reliably construct the matrix element
$\langle \pi^+\pi^-|Q_6|K^0\rangle$ within the quenched
approximation using quenched ChPT and our lattice data.  This is
due to the large value of $\alpha_q^{NS}$, a low energy constant
absent outside of the quenched approximation, which was implicitly
assumed to be zero in the previous RBC and CP-PACS work. Such a
large quenching artifact does not merely make the extraction of
$\langle \pi^+\pi^-|Q_6|K^0\rangle$ practically difficult, but
implies that the quenched approximation itself fails to accurately
describe the full theory.  In fact, the motivation for the
definition of the quenched approximation outlined in Section II is
seen to be invalid, given the large differences in the analytic
structure of the full and quenched theories.  This clearly reduces
the physical relevance of such a calculation.

These arguments and results demonstrate that the errors associated
with the quenched approximation in the evaluation of the strong
penguin operators are likely to be quite significant.  Definitive
answers will clearly have to await dynamical simulations.  The
first steps in performing such full QCD simulations have been
underway for quite some time with two dynamical
flavors~\cite{Aoki:2004ht}; the more realistic 2+1 flavor
simulations have begun.

\centerline{\bf ACKNOWLEDGEMENTS}

We thank Maarten Golterman for discussions, our RBC colleagues and
especially Thomas Blum and Robert Mawhinney for helpful
discussions and suggestions.  This research was supported in part
by the US DOE under Contract No. DE-AC02-98CH10886 (Laiho, Soni),
in part by the DOE under grant \# DE-FG02-92ER40699 (Columbia), in
part by the RIKEN-BNL Research Center (Blum, Dawson, Noaki), in
part by the LDRD grant (Blum, Laiho), and in part by the DOE under
grant \# DE-AC02-76CH03000 (Laiho).

\appendix
\section*{\bf APPENDIX A}
\setcounter{equation}{0} \setcounter{section}{1}
\renewcommand{\theequation}{A\arabic{equation}}

The function, $F$, appearing in the one-loop contribution to
$K\to\pi\pi$ is given by \cite{Golterman:2002us}

\bea  F(m^2_1,m^2_2,p^2) &=&
\sqrt{\lambda\left(1,\frac{m^2_1}{p^2},\frac{m^2_2}{p^2}\right)}
\ln\frac{p^2+m^2_1+m^2_2+p^2\sqrt{\lambda(1,m^2_1/p^2,m^2_2/p^2)}}
{p^2+m^2_1+m^2_2-p^2\sqrt{\lambda(1,m^2_1/p^2,m^2_2/p^2)}},
\nonumber \\ && \eea

with

\bea \lambda(x,y,z)=(x-y+z)^2+4xy. \nonumber  \eea

\appendix
\section*{\bf APPENDIX B: Normalization Conventions}
\setcounter{equation}{0} \setcounter{section}{1}
\renewcommand{\theequation}{B\arabic{equation}}

In this appendix we specify the sign and normalization conventions
used in this paper.  This includes both the fields and operators
used in chiral perturbation theory (where we follow in most cases
the conventions used in the earlier work of Golterman and
Pallante) and the corresponding quantities defined on the quark
level where we follow the conventions used in
Ref.~\cite{Blum:2001xb}.

Equations~(23) and (24) define the relative signs of the pseudo
scalar fields $\phi$, the pseudo-fermion fields $\chi$ and the
ghost fields $\widetilde{\phi}$ in the sense that a specific pair
of $SU(3|3)_L \times SU(3|3)_R$ matrices, $(U_L, U_R)$ will
transform these fields in a determined fashion:
\begin{equation}
\Sigma \rightarrow \Sigma^\prime = U_L \Sigma U_R^\dagger.
\label{eq:trans_chiral}
\end{equation}
Note this relation between the fields $\phi$, $\chi$ and
$\widetilde{\phi}$ is still somewhat abstract because the $6
\times 6$ matrices $U_L$ and $U_R$ contain both commuting and
anti-commuting numbers. However, this description will be
sufficient for our purposes because these same matrices transform
the quark and pseudo-quark fields $q$ and $\widetilde{q}$:
\begin{equation}
{\cal Q}\equiv\left( \begin{array}{c} q \\ \widetilde{q} \\
\end{array} \right) \rightarrow \left( \begin{array}{c} q^\prime
\\ \widetilde{q}^\prime \\ \end{array} \right) = U_R \left(
\begin{array}{c} q_R \\ \widetilde{q}_R \\ \end{array} \right) +
U_L \left( \begin{array}{c} q_L \\ \widetilde{q}_L \\ \end{array}
\right) = \Bigl(U_R P_R + U_L P_L\Bigr){\cal Q}.
\label{eq:trans_quark}
\end{equation}
Here $P_{R/L}=(1\pm \gamma^5)/2$ and as in the text, ${\cal Q}$
contains the three flavors of quarks $q$ and the three flavors of
pseudo quarks $\widetilde{q}$ and belongs to the Cartesian product
representation $SU(3|3)_L \times SU(3|3)_R$.

The absolute normalization and sign for the $\Sigma$ field is
determined by the lowest order effective chiral Lagrangian for
quenched QCD:
\begin{equation}
{\cal L}^{(2)}_{QCD} = \frac{f^2}{8}{\rm str}\bigl\{\partial_\mu
\Sigma^\dagger \partial^\mu \Sigma\bigr\}
                 + \frac{B_0 f^2}{4} \,
                 {\rm str}\bigl\{M^\dagger \Sigma + \Sigma^\dagger
                 M\bigr\}
\end{equation}
(written in Minkowski-space) once we specify that the parameter
$B_0$ is real and positive with a magnitude chosen so that the $6
\times 6$ matrix $M$ is the quark mass matrix appearing in the
fundamental QCD Lagrangian written below.

The Dirac fields $q$ and $\widetilde{q}$ are conventional, with
the Minkowski space Lagrangian given in terms of ${\cal Q}$ by
\begin{equation}
    {\cal L} = \overline{{\cal Q}}i\gamma^\mu D_\mu {\cal Q}
       - \overline{{\cal Q}}(M^\dagger P_L + M P_R)
        {\cal Q}\ .
\end{equation}
The connection between quark fields and the corresponding
quantities in the effective chiral Lagrangian is given by the
equation:
\begin{equation}
    ({\cal Q}_L)_j (\overline{{\cal Q}}_R)_i =
    \frac{B_0 f^2}{4}\, \Sigma_{j,i}.
\label{eq:ps_conv}
\end{equation}
This equation is a generalization of the usual relation between
$\Sigma$ and the quark fields to include the new variables which
appear in the quenched case.  It is uniquely determined by the
combined requirements of flavor covariance (the left and
right-hand sides transform identically under the $SU(3|3)_L \times
SU(3|3)_R$ flavor transformations of Eqs.~(\ref{eq:trans_chiral})
and (\ref{eq:trans_quark})) and consistency with the equation for
the chiral condensate:
\begin{equation}
\langle \overline{u_R} u_L \rangle
    = \frac{i\partial}{\partial M^\dagger_{11}}\ln
    Z[M,M^\dagger]
    = -\frac{B_0 f^2}{4} \langle\Sigma_{11}\rangle,
\end{equation}
where we examine the condensate associated with the up quark, the
first component of ${\cal Q}$, and treat the matrix elements
$M_{1,1}$ and $M_{1,1}^*$ as independent variables.  Note, the
flavor covariance of Eq.~(\ref{eq:ps_conv}) is lost and the
resulting equation invalid if the order of the factors $({\cal
Q}_R)_j$ and $(\overline{{\cal Q}}_L)_i$ is reversed, since these
6-component fields are a mixture of commuting and anti-commuting
quantities.

Following standard conventions, we identify the field $\phi$ in
Eq.~(24) with meson fields according to:
\begin{equation}
\phi =
    \left(\begin{array}{ccc}
       \pi^0/\sqrt{2}+\eta/\sqrt{6} & \pi^+  & K^+ \\
       \pi^-  & -\pi^0/\sqrt{2}+\eta/\sqrt{6} & K^0 \\
       K^-  & \overline{K^0}  & -2\eta/\sqrt{6}
    \end{array}\right),
\label{eq:meson_conv}
\end{equation}
where the specific fields appearing in Eq.~(\ref{eq:meson_conv})
above destroy the corresponding mesons.  Similarly the ghost field
$\chi$ of Eq. 24 can be written in an identical fashion if a tilde
is added to each of the meson fields, {\it e.g.} $\pi^+$ is
replaced by $\widetilde{\pi}^+$. With these conventions we can
then examine specific components of Eq.~(\ref{eq:ps_conv}) to
lowest order in chiral perturbation theory and obtain the useful
relations:
\begin{eqnarray}
    \overline{s}\gamma^5 d &=& iB_0 f
     K^0 \quad (j=2, i=3)\ , \\
    \overline{s}\gamma^5 \widetilde{d} &=&
    -iB_0 f \widetilde{K}^0 \quad (j=5, i=3).
\end{eqnarray}
This same identification can be made by an appropriate
generalization of the results in Appendix A of
Ref.~\cite{Blum:2001xb} to the quenched case.

Remaining consistent with Eq.~(\ref{eq:ps_conv}), we can easily
write down expressions relating quark-level operator states and
chiral-level operator states. For a given meson operator in terms
of the underlying quark fields, we have
\begin{eqnarray}
  \overline{d}\gamma_5 s |0\rangle & = &
   iB_0 f \overline{K}^0 |0\rangle  = iB_0f|K^0\rangle
    \label{eq:K0state}\ ,\\
  \overline{\widetilde{d}}\gamma_5 s |0\rangle & = &
  - iB_0 f (\widetilde{K}^0)^\dagger
   |0\rangle
   = -iB_0f|\tilde{K}^0\rangle
   \label{eq:K0tildestate}\ ,
\end{eqnarray}
where Eq.~(\ref{eq:K0state}) is the standard relation between
states in terms of the quark-level and chiral operators.
Eq.~(\ref{eq:K0tildestate}) is found by a chiral rotation on
Eq.~(\ref{eq:K0state}) to give the corresponding relation between
these fermionic meson states. To be concrete, we also include here
the rules needed for evaluating matrix elements of operators with
these $\chi$ fields. These are given by chiral rotations on the
appropriate $K^0$ field rules, and found to be
\begin{equation}\label{eq:K0tildeME}
    \langle 0 | \widetilde{K}^0(x) |\widetilde{K}^0(k)\rangle
    = e^{ikx}\ ,
\end{equation}
and
\begin{equation}
    \langle \widetilde{K}^0(\mathbf{k}') |
    \widetilde{K}^0(\mathbf{k})\rangle
    = 2 E_K (2\pi)^3 \delta^3(\mathbf{k}' -\mathbf{k}) \ .
\end{equation}
Unlike the standard meson fields like $K^0$, the ordering in these
expressions is rather important, since $\widetilde{K}^0
(\widetilde{K}^0)^\dagger =-(\widetilde{K}^0)^\dagger
\widetilde{K}^0 $. Additionally, as stated earlier in the text,
the quark and ghost-quark propagators are equal flavor by flavor
(when properly ordered), such that
\begin{equation}
 \langle 0 | d(x) \overline {d}(y)| 0 \rangle
   = \langle 0 | \widetilde{d}(x) \overline{\widetilde{d}}(y)
   | 0 \rangle
   =  S_F(x,y)\ ,
\end{equation}
where $(\gamma_\mu D_\mu +m )S_F(x,y) = \delta^{(4)}(x-y)$, for
the case of continuum, Euclidean fermions.

Next we apply this same approach to relate the various components
of the four-quark operator related to $Q_6^{QNS}$ as they appear
in chiral perturbation theory and at the quark level:
\begin{equation}\label{eq:gen_QNS_op}
    {\cal O}^{QNS}_{ji} \equiv
    {\rm tr}_D\left\{(Q_L)_j^a
    (\gamma^\mu)^t(\overline{{\cal Q}}_L)_i^b \right\}\,
      {\rm tr}_D\left\{{\rm str}\left[{\cal Q}_R^b (\gamma^\mu)^t
      \overline{{\cal Q}}_R^a
      \overline{A}\right]\right\}
    = \alpha^{NS}_q\Bigl(\Sigma\overline{A}
    \Sigma^\dagger\Bigr)_{j,i}
\end{equation}
where the diagonal matrix $\overline{A}$ is defined in Eq.~(21),
$a$ and $b$ are color indices, ${\rm tr}_D$ is a trace over the
(implied) Dirac indices, and we are exploiting flavor covariance
and the conventions of Ref.~\cite{Golterman:2003yw}. We can use
this expression to evaluate two cases of importance by using
$(\lambda_6)_{ij} = \delta_{i3}\delta_{j2}$ and
$(\widetilde\lambda_6)_{ij} = \delta_{i3}\delta_{j5}$. In other
words, multiplying Eq.~(\ref{eq:gen_QNS_op}) by either $\lambda_6$
or $\widetilde\lambda_6$ and taking the supertrace over the
$SU(3|3)$ indices, we get
\begin{eqnarray}
    Q_6^{QNS} & = & {\rm str}\left[ \lambda_6 {\cal O}^{QNS}\right]
    + {\rm h.c.}
    = \alpha^{NS}_q
   {\rm str}\bigl(\lambda_6
   \Sigma\overline{A}\Sigma^\dagger\bigr) + {\rm h.c.}
    \ ,\\
    \widetilde{Q}_6^{QNS} & = &
    {\rm str}\left[ \widetilde\lambda_6 {\cal O}^{QNS}\right]
    + {\rm h.c.}
    = \alpha^{NS}_q
    {\rm str}\bigl(\widetilde\lambda_6
    \Sigma\overline{A}\Sigma^\dagger\bigr)
    + {\rm h.c.} \ ,
\end{eqnarray}
which are precisely the expressions we see in Eqs.~(14) and (19)
for $Q_6^{QNS}$, and Eqs.~(42) and (43) for
$\widetilde{Q}_6^{QNS}$.

\appendix
\section*{\bf APPENDIX C}
\setcounter{equation}{0} \setcounter{section}{1}
\renewcommand{\theequation}{C\arabic{equation}}

The NLO corrections to $K\to\widetilde{\pi}$ are calculated from
one-loop insertions of the operator Eq.~\ref{26.5}, as well as
local operators that begin at NLO:

\bea
\widetilde{{\cal O}}^{NS}_{1}= \textrm{tr}[\widetilde{\lambda_{6}} L_{\mu}\Sigma^{\dag} \overline{A} \Sigma L^{\mu}],\\
 \widetilde{{\cal O}}^{NS}_{2}= \textrm{tr}[\widetilde{\lambda_{6}} L_{\mu}]\textrm{tr}[\Sigma^{\dag} \overline{A} \Sigma L^{\mu}],\\
 \widetilde{{\cal O}}^{NS}_{3}= \textrm{tr}[\widetilde{\lambda_{6}} \{\Sigma^{\dag} \overline{A} \Sigma, L^{2}\}],\\
 \widetilde{{\cal O}}^{NS}_{4}= \textrm{tr}[\widetilde{\lambda_{6}} \{\Sigma^{\dag} \overline{A} \Sigma, S\}],\\
 \widetilde{{\cal O}}^{NS}_{5}= \textrm{tr}[\widetilde{\lambda_{6}} [\Sigma^{\dag} \overline{A} \Sigma, P]],\\
 \widetilde{{\cal O}}^{NS}_{6}= \textrm{tr}[\widetilde{\lambda_{6}} \Sigma^{\dag} \overline{A} \Sigma]
 \textrm{tr}[S].\eea

\noindent Each operator has associated with it an a priori unknown
LEC which we call $\tilde{c}_i^{NS}$, with a scale dependence
similar to that of Eq~\ref{16},

\begin{equation}\label{B}
\tilde{c}^{r,NS}_i(\mu_2)=\tilde{c}^{r,NS}_i(\mu_1)+\frac{2\alpha_{q}^{NS}\widetilde{\eta}_i}{(4\pi
f)^2}\ln{\frac{\mu_1}{\mu_2}}.
\end{equation}

The coefficient of the scale dependence, $\widetilde{\eta}_i$, can
be determined as in \cite{Cirigliano:1999pv}, where the authors
apply background field and heat bath methods.  The values for
$\widetilde{\eta}_i$ are $0$, $-2$, $-N_f/2$, $N_f/2$, $0$ and 1
for $i=1-6$ respectively. In this case $N_f$ is the number of sea
quarks, which for the quenched case should naively be taken to
zero. However one must take care in this case because this
operator (Eq~\ref{26.5}) has been introduced precisely to add
contractions of the four quark operators that contribute to the
fermion determinant which would otherwise be absent in the
quenched approximation in which we are working.

We can determine the scale dependence in the quenched theory by
making use of the following trick. In the partially quenched case,
$\overline{A}\to\overline{A}_{PQ}=2(1-3/N_f,1-3/N_f,1-3/N_f,-3/N_f,...,-3/N_f)_{\rm
diag}$, where the first 3 valence entries are $1-3/N_f$, and the
next $N_f+3$ ghost and sea entries are $-3/N_f$.  Taking the
$N_f\to 0$ limit in this matrix is singular, but when these
factors multiply the $\widetilde{\eta}_i$ factors above, this
limit yields the correct scale dependence for the amplitudes in
the quenched theory.

The scale dependence obtained in this way agrees with that of a
direct one-loop calculation of $K\to\widetilde{\pi}$, which yields
precisely zero in the case of degenerate quarks.  Thus, the
one-loop chiral log vanishes, and the only NLO contribution in the
degenerate case is proportional to $m_M^2$ ($m_f+m_{res}$ in fits
to DW fermions) with an unknown coefficient.  The diagrams needed
are given in Fig.~\ref{feyn}. The wave-function renormalization
vanishes for $\widetilde{\pi}$ and $K$ for this matrix element.

\bibliography{paper}

\newpage


\begin{table}
\caption{The ratio $(\langle 0|\tilde{Q}^{QNS}_6|\tilde{K}\rangle
- \langle 0|Q^{QNS}_6|K\rangle)/\langle 0|\overline{s}\gamma_5
d|K\rangle$ for each of ten nondegenerate pairs of quark masses.
The chiral limit of this ratio ($m_d \rightarrow 0$) can be used
to obtain $\alpha_q^{NS}$.} \label{tabone}
\small{\begin{tabular}{ccccc} \hline \hline
  $m_s$ & $m_d=0.01$ & $m_d=0.02$ & $m_d=0.03$ & $m_d=0.04$  \\
  \hline
  0.02 & $-2.898(13)\times 10^{-2}$ &  & &  \\
  0.03 & $-2.918(13)\times 10^{-2}$ & $-5.010(13)\times 10^{-2}$ & & \\
  0.04 & $-2.932(13)\times 10^{-2}$ & $-5.020(12)\times 10^{-2}$& $-7.088(13)\times 10^{-2}$ & \\
  0.05 & $-2.941(13)\times 10^{-2}$ & $-5.026(12)\times 10^{-2}$& $-7.093(13)\times 10^{-2}$
   & $-9.141(14)\times 10^{-2}$ \\
\hline \hline
\end{tabular}}
\end{table}

\begin{table}
\caption{The matrix element $\langle
\tilde{\pi}|\tilde{Q}^{QNS}_6|K\rangle$ as a function of $m_f$,
the single quark mass that appears in the degenerate kaon and pion
states.  The chiral limit of this data can be used to obtain
$\alpha_q^{NS}$ using Eq.~\ref{43.5}.}
 \label{tabtwo}
\begin{tabular}{cc}
\hline \hline
  $m_f$ & $\langle \tilde{\pi}|\tilde{Q}^{QNS}_6|K\rangle$  \\
  \hline
  0.01 &  $1.702(69)\times 10^{-2}$   \\
  0.02 &  $2.293(68)\times 10^{-2}$  \\
  0.03 &  $2.910(70)\times 10^{-2}$  \\
  0.04 &  $3.545(72)\times 10^{-2}$  \\
  0.05 &  $4.199(74)\times 10^{-2}$  \\
\hline \hline
\end{tabular}
\end{table}

\clearpage


\begin{figure}
\vskip -1.5in
\centering
\begin{fmffile}{vp_sd}
  \begin{fmfchar*}(80,120)
    \fmfleft{sq} \fmflabel{\large $s$}{sq}
    \fmf{fermion,tension=4.0}{Wst,sq}
    \fmf{fermion,tension=1.2}{tt2,Wst}
    \fmf{gluon, label={\large $g$},tension=0.4}{tt2,vp2}
    \fmf{fermion,label={\large $q$},label.side=right,label.dist=5,right=0.8,tension=0.2}{vp1,vp2}
    \fmf{fermion,label={\large $q$},label.side=right,label.dist=5,right=0.8,tension=0.2}{vp2,vp1}
    \fmf{gluon, label={\large $g$},tension=0.4}{tt1,vp1}
    \fmf{photon, left=0.5,label.side=right,label=$W$,tension=1.2}{Wst,Wtd}
    \fmf{fermion,label={\large $t$},label.side=left,label.dist=5,tension=0.4}{gtt,tt2}
    \fmf{fermion,label={\large $t$},label.side=left,label.dist=4,tension=0.4}{tt1,gtt}
    \fmf{fermion,tension=4.0}{dq,Wtd}
    \fmf{fermion,tension=1.2}{Wtd,tt1}
    \fmf{gluon, label={\large $g$},tension=1.0}{gtt,guu}
    \fmfright{dq} \fmflabel{\large $d$}{dq}
    \fmf{fermion}{ui,guu}\fmf{fermion}{guu,uo}
    \fmfbottom{ui,uo} \fmflabel{\large $u$}{ui} \fmflabel{\large $u$}{uo}
    \fmfdot{Wst,Wtd,gtt,guu}
  \end{fmfchar*}
\end{fmffile}
\caption{A $K^+ \rightarrow \pi^+$ diagram in which the vacuum polarization
loop is entirely contained within a high-momentum subgraph.}
\label{fig:vac_pol_sd}
\end{figure}
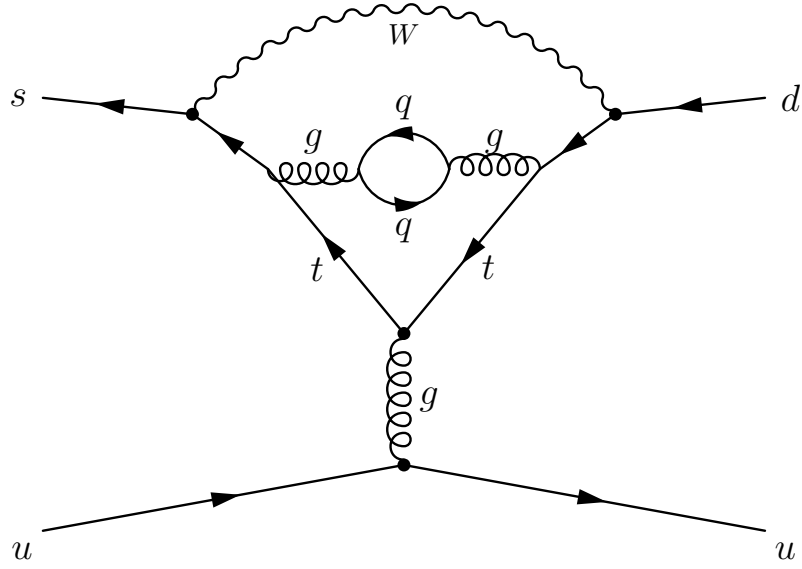

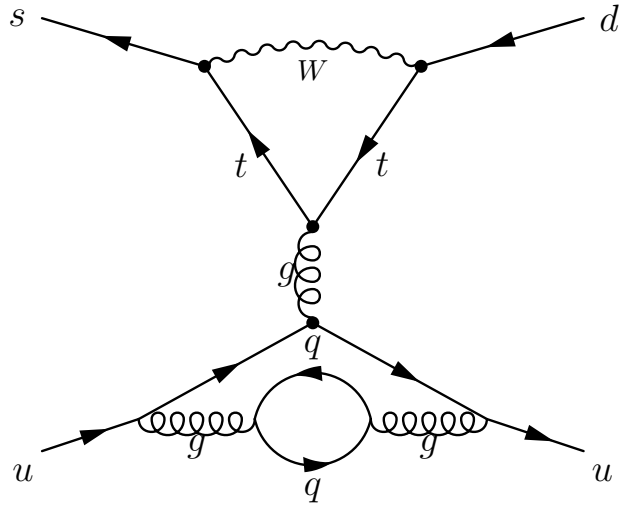
\begin{figure}
\centering
\begin{fmffile}{vp_ld}
  \begin{fmfchar*}(60,120)
    \fmfleft{sq} \fmflabel{\large $s$}{sq}
    \fmf{fermion,tension=2.0}{Wst,sq}
    \fmf{photon, left=0.2,label.side=right,label=$W$,tension=1.2}{Wst,Wtd}
    \fmf{fermion,label={\large $t$},label.side=left,label.dist=5,tension=0.6}{gtt,Wst}
    \fmf{fermion,label={\large $t$},label.side=left,label.dist=4,tension=0.6}{Wtd,gtt}
    \fmf{fermion,tension=2.0}{dq,Wtd}
    \fmf{gluon,tension=2.0, label={\large $g$}}{gtt,guu}
    \fmfright{dq} \fmflabel{\large $d$}{dq}
    \fmf{fermion,tension=3.0}{ui,uu1}
    \fmf{fermion}{uu1,guu}
    \fmf{gluon, label={\large $g$}}{uu1,vp1}
    \fmf{fermion,label={\large $q$},label.side=right,label.dist=5,right=0.8,tension=0.5}{vp2,vp1}
    \fmf{fermion,label={\large $q$},label.side=right,label.dist=5,right=0.8,tension=0.5}{vp1,vp2}
    \fmf{gluon, label={\large $g$}}{vp2,uu2}
    \fmf{fermion}{guu,uu2}
    \fmf{fermion,tension=3.0}{uu2,uo}
    \fmfbottom{ui,uo} \fmflabel{\large $u$}{ui} \fmflabel{\large $u$}{uo}
    \fmfdot{Wst,Wtd,gtt,guu}
  \end{fmfchar*}
\end{fmffile}
\caption{A $K^+ \rightarrow \pi^+$ diagram in which the vacuum polarization
subgraph must appear entirely within a low-momentum subgraph.}
\label{fig:vac_pol_ld}
\end{figure}

\begin{figure}
\centering
\begin{fmffile}{vp_sold}
\begin{fmfchar*}(60,120)
  \fmfleft{sq} \fmflabel{\large $s$}{sq}
    \fmf{fermion,tension=1.5}{gss,sq}
    \fmf{fermion}{Wsu,gss}
    \fmf{fermion,tension=1.5}{uqb,Wsu}
  \fmfright{uqb} \fmflabel{\large $u$}{uqb}
    \fmf{gluon, label={\large $g$}}{gss,gq1q1}
      \fmf{fermion,label={\large $q$},label.side=right,label.dist=5,right=0.8,tension=0.5}{gq1q1,gq2q2}
      \fmf{fermion,label={\large $q$},label.side=right,label.dist=5,right=0.8,tension=0.5}{gq2q2,gq1q1}
    \fmf{gluon, label={\large $g$}}{gq2q2,guu}
    \fmf{photon,label=$W$,tension=0.35}{Wsu,Wud}
  \fmfleft{uq} \fmflabel{\large $u$}{uq}
    \fmf{fermion,tension=1.5}{uq,guu}
    \fmf{fermion}{guu,Wud}
    \fmf{fermion,tension=1.5}{Wud,dq}
  \fmfright{dq} \fmflabel{\large $d$}{dq}
  \fmfbottom{uq,dq}
  \fmfdot{Wsu,Wud,gss,guu}
\begin{fmfsubgraph}(20,20)(160,200)
  \fmftop{t1,t2}
  \fmf{dashes}{t1,t2}
  \fmf{dashes}{t2,b2}
  \fmf{dashes}{b2,b1}
  \fmf{dashes}{b1,t1}
  \fmfbottom{b1,b2}
\end{fmfsubgraph}
\begin{fmfsubgraph}(100,30)(70,180)
  \fmftop{t1a,t2a}
  \fmf{dashes}{t1a,t2a}
  \fmf{dashes}{t2a,b2a}
  \fmf{dashes}{b2a,b1a}
  \fmf{dashes}{b1a,t1a}
  \fmfbottom{b1a,b2a}
\end{fmfsubgraph}

\end{fmfchar*}
\end{fmffile}
\caption{A $K^+ \rightarrow \pi^+$ diagram in which the vacuum polarization
subgraph must be wholly inside or outside of any high-momentum subgraph.}
\label{fig:vac_pol_sd_or_ld}
\end{figure}
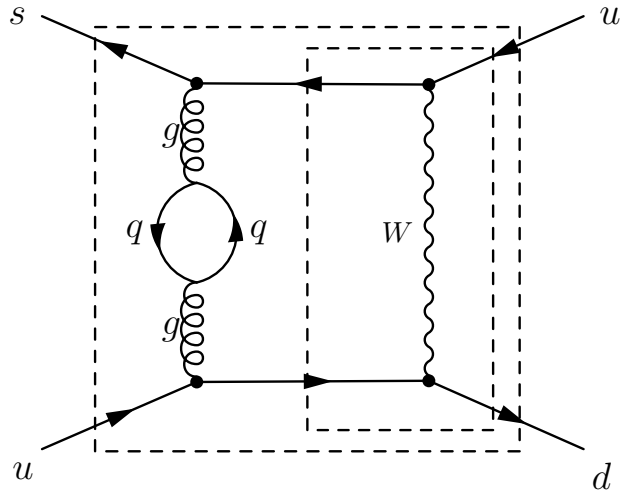

\begin{figure}
\centering
\begin{fmffile}{vp_sald}
  \begin{fmfchar*}(60,120)
    \fmfleft{sq} \fmflabel{\large $s$}{sq}
    \fmf{fermion,tension=2.0}{Wst,sq}
    \fmf{photon,left=0.5,label.side=right,label=$W$,tension=1.2}{Wst,Wtd}
    \fmf{fermion,label={\large $t$},label.side=left,label.dist=5,tension=0.6}{gtt,Wst}
    \fmf{fermion,label={\large $t$},label.side=left,label.dist=4,tension=0.6}{Wtd,gtt}
    \fmf{fermion,tension=2.0}{dq,Wtd}
    \fmf{gluon, label={\large $g$}}{gtt,gq1q1}
    \fmfright{dq} \fmflabel{\large $d$}{dq}
    \fmf{fermion,label={\large $q$},label.side=right,label.dist=5,right=0.8,tension=0.4}{gq1q1,gq2q2}
    \fmf{fermion,label={\large $q$},label.side=right,label.dist=5,right=0.8,tension=0.4}{gq2q2,gq1q1}
    \fmf{gluon, label={\large $g$}}{gq2q2,guu}
    \fmf{fermion}{ui,guu}\fmf{fermion}{guu,uo}
    \fmfbottom{ui,uo} \fmflabel{\large $u$}{ui} \fmflabel{\large $u$}{uo}
    \fmfdot{Wst,Wtd,gtt,guu}

\begin{fmfsubgraph}(30,10)(150,260)
  \fmftop{t1,t2}
  \fmf{dashes}{t1,t2}
  \fmf{dashes}{t2,b2}
  \fmf{dashes}{b2,b1}
  \fmf{dashes}{b1,t1}
  \fmfbottom{b1,b2}
\end{fmfsubgraph}
\begin{fmfsubgraph}(40,95)(130,155)
  \fmftop{t1a,t2a}
  \fmf{dashes}{t1a,t2a}
  \fmf{dashes}{t2a,b2a}
  \fmf{dashes}{b2a,b1a}
  \fmf{dashes}{b1a,t1a}
  \fmfbottom{b1a,b2a}
\end{fmfsubgraph}
\begin{fmfsubgraph}(50,140)(110,90)
  \fmftop{t1b,t2b}
  \fmf{dashes}{t1b,t2b}
  \fmf{dashes}{t2b,b2b}
  \fmf{dashes}{b2b,b1b}
  \fmf{dashes}{b1b,t1b}
  \fmfbottom{b1b,b2b}
\end{fmfsubgraph}

\end{fmfchar*}
\end{fmffile}
\caption{A $K^+ \rightarrow \pi^+$ diagram in which the vacuum polarization
subgraph can be partially contained within a high-momentum subgraph.}
\label{fig:vac_pol_sd_and_ld}
\end{figure}
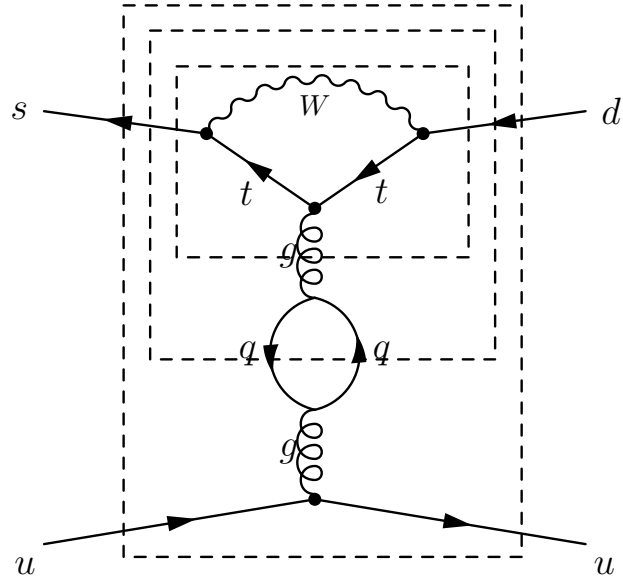

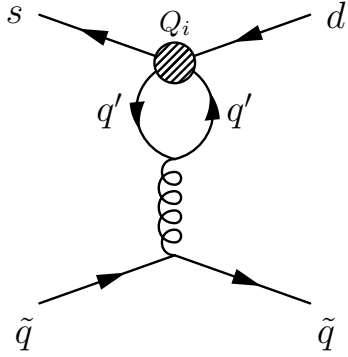
\begin{figure}
\begin{fmffile}{gh_gen}
  \begin{fmfchar*}(30,80)
    \fmfleft{dq} \fmflabel{\large $s$}{dq}
    \fmf{fermion}{Qi,dq}
    \fmf{fermion}{sq,Qi}
    \fmfright{sq} \fmflabel{\large $d$}{sq}
    \fmfv{d.sh=circle,d.f=shaded,d.si=0.15w,l.d=10,l.a=90,l=$Q_i$}{Qi}
    \fmf{fermion,label={\large $q'$},label.side=right,label.dist=5,right=0.8,tension=0.5}{Qi,gqq}
    \fmf{fermion,label={\large $q'$},label.side=right,label.dist=5,right=0.8,tension=0.5}{gqq,Qi}
    \fmf{gluon}{gqq,gq1q1}
    \fmf{fermion}{qi,gq1q1}\fmf{fermion}{gq1q1,qo}
    \fmfbottom{qi,qo} \fmflabel{\large $\tilde{q}$}{qi} \fmflabel{\large $\tilde{q}$}{qo}
  \end{fmfchar*}
\end{fmffile}
\caption{An order $\alpha_s$ diagram which generates a non-vanishing ghost amplitude in a
quenched theory from operators with no direct ghost quark coupling.}
\label{fig:induced_ghosts}
\end{figure}

\begin{figure}
\begin{center}
\includegraphics[scale=.7]{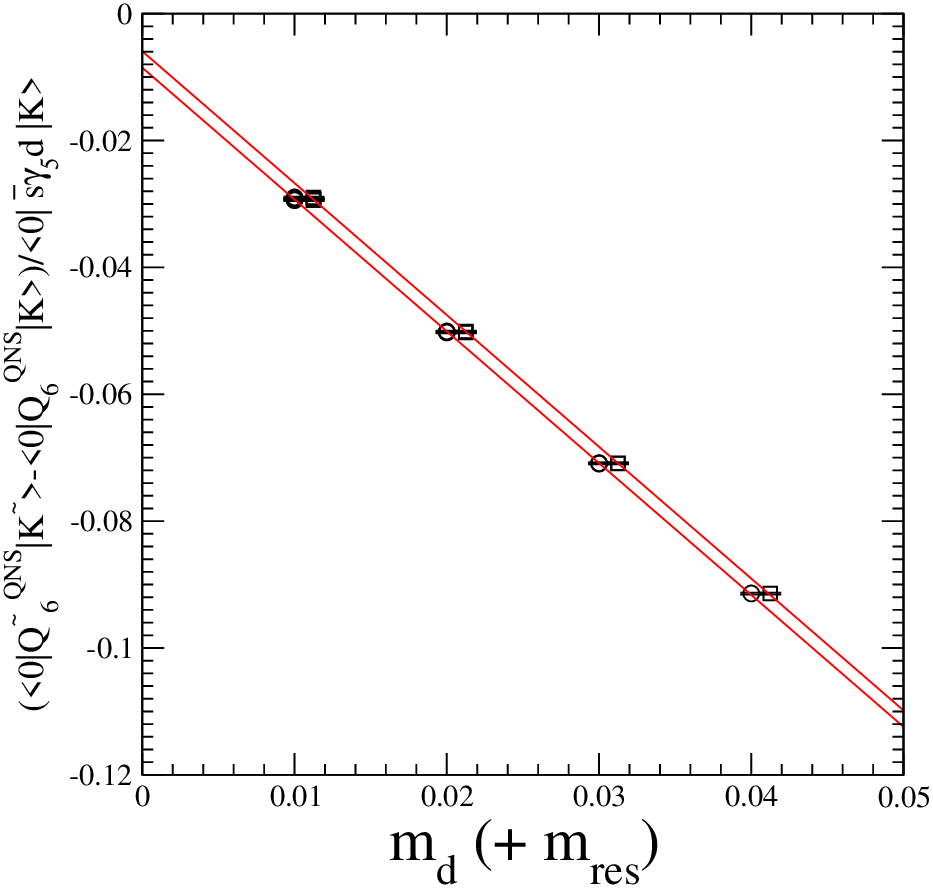}
\end{center}

\caption{The ratio $(\langle 0|\tilde{Q}^{QNS}_6|\tilde{K}\rangle
- \langle 0|Q^{QNS}_6|K\rangle)/\langle 0|\overline{s}\gamma_5
d|K\rangle$ versus $m_d$ (circles) and $m_d+m_{res}$ (squares) for
the ten nondegenerate quark masses. The lines are linear fits to
the data of the form $\eta_0+\eta_1 m_d$ and
$\eta_0+\eta_1(m_d+m_{res})$. \label{fig6}}
\end{figure}

\begin{figure}
\begin{center}
\includegraphics[scale=.6]{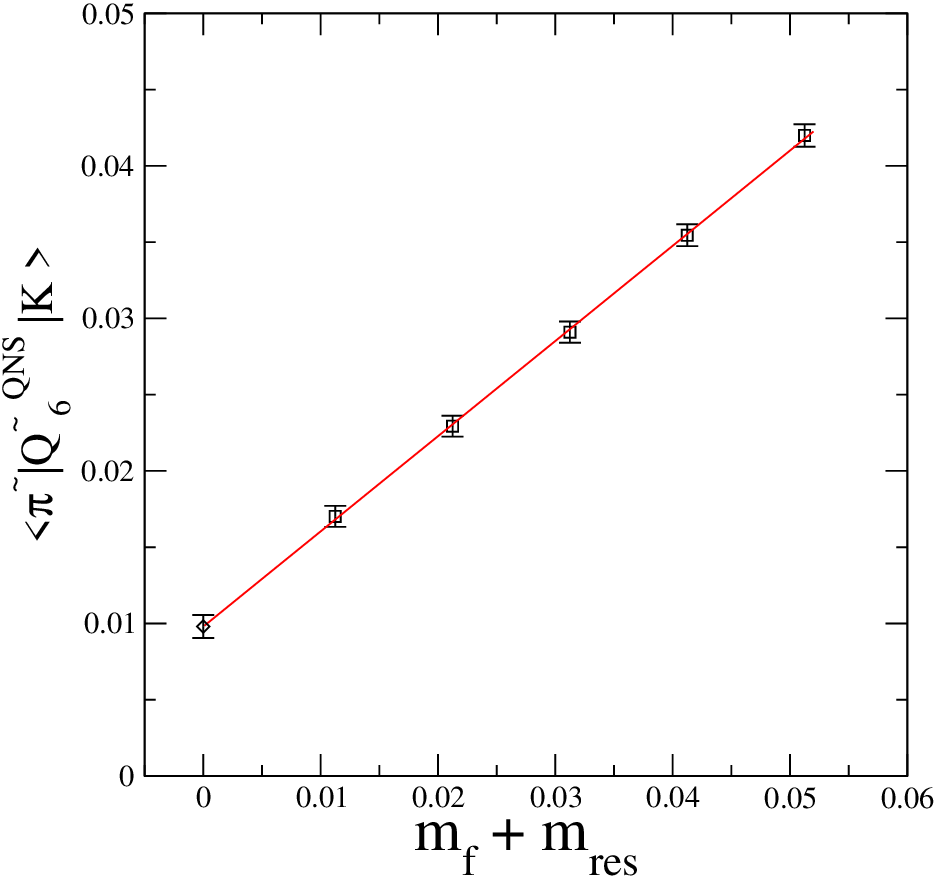}
\end{center}
\caption{The matrix element $\langle
\tilde{\pi}|\tilde{Q}^{QNS}_6|K\rangle$ as a function of
$m_f+m_{res}$.  The solid line is a simple linear fit to the data
and the intercept can be used to obtain $\alpha_q^{NS}$ from
Eq.~\ref{43.5}.   \label{fig7}}
\end{figure}

\begin{figure}
\begin{center}
\includegraphics[scale=.8]{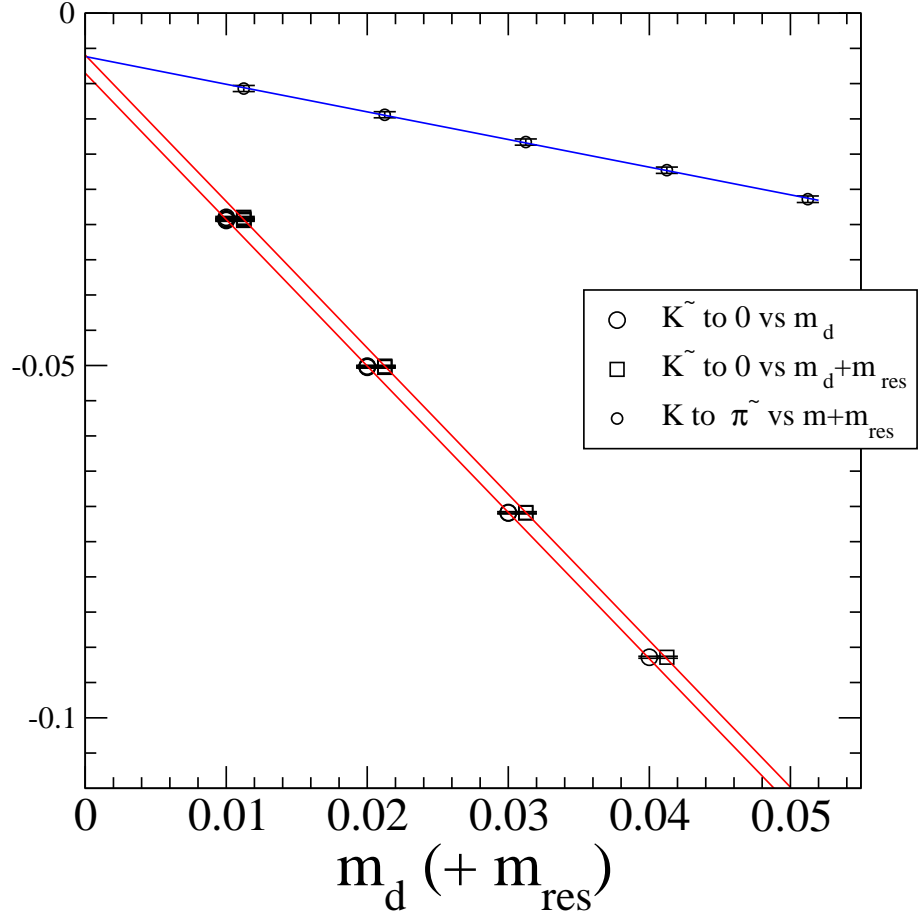}
\end{center}
\caption{The matrix elements $(\langle
0|\tilde{Q}^{QNS}_6|\tilde{K}\rangle - \langle
0|Q^{QNS}_6|K\rangle)/\langle 0|\overline{s}\gamma_5 d|K\rangle$
as a function of $m_d$ (big circles) and $m_d+m_{res}$ (squares)
and $f^2/(2\alpha^{(3,\overline{3})})\langle
\tilde{\pi}|\tilde{Q}^{QNS}_6|K\rangle$ (small circles) as a
function of $m_d$ equal to the degenerate masses of the quarks
appearing in the $K$ meson.  For comparison, we plot the
$\widetilde{K}\to 0$ matrix elements above. The pre-factor of
$\langle \tilde{\pi}|\tilde{Q}^{QNS}_6|K\rangle$ re-scales the
matrix element so that the chiral limits of the two methods are
expected to agree according to ChPT.  This graph shows that the
chiral limits do agree, thus yielding similar values of
$\alpha_q^{NS}$. \label{fig8}}
\end{figure}

\begin{figure}[htbp]
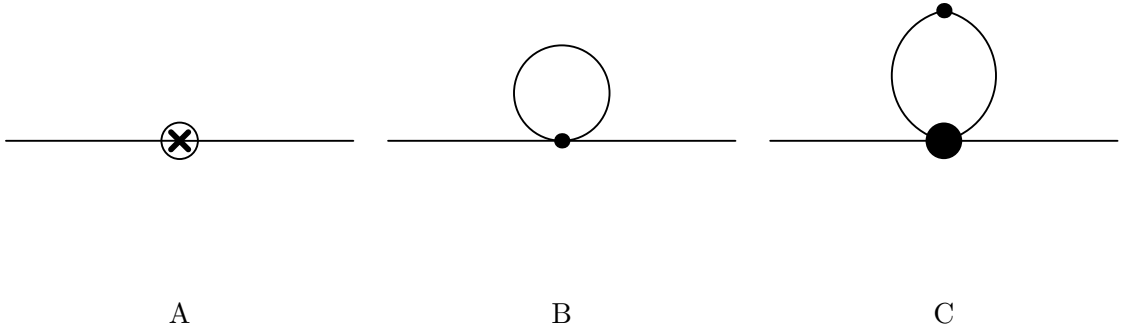


\unitlength=1bp%

\begin{feynartspicture}(432,445)(3,1)

\FADiagram{A} \FAProp(0.,10.)(10.,10.)(0.,){/Straight}{0}
\FAProp(20.,10.)(10.,10.)(0.,){/Straight}{0} \FAVert(10.,10.){2}

\FADiagram{B} \FAProp(0.,10.)(10.,10.)(0.,){/Straight}{0}
\FAProp(20.,10.)(10.,10.)(0.,){/Straight}{0}
\FAProp(10.,10.)(10.,10.)(10.,15.5){/Straight}{0}
\FAVert(10.,10.){0}

\FADiagram{C} \FAProp(0.,10.)(10.,10.)(0.,){/Straight}{0}
\FAProp(20.,10.)(10.,10.)(0.,){/Straight}{0}
\FAProp(10.,17.5)(10.,10.)(-0.8,){/Straight}{0}
\FAProp(10.,17.5)(10.,10.)(0.8,){/Straight}{0}
\FAVert(10.,10.){-5} \FAVert(10.,17.5){0}

\end{feynartspicture}

\caption{Diagrams needed to evaluate the NLO amplitude $K\to
\widetilde{\pi}$. NLO corrections include tree-level diagrams with
insertion of the NLO weak vertices (crossed circle), one-loop
diagrams with insertions of the LO weak vertices (small filled
circles) and the $O(p^{2})$ strong vertices (big filled circle).
The lines represent the propagators of mesons comprised of valence
and ghost quarks. \label{feyn}}
\end{figure}

\end{document}